\documentclass[aps, prb, floatfix, superscriptaddress, twocolumn]{revtex4-1}

\usepackage{graphicx}
\usepackage{color}
\usepackage{calc}
\usepackage{array}
\usepackage{graphicx}
\usepackage{amsmath, amssymb}
\usepackage{natbib}
\usepackage{hyperref}

\DeclareMathOperator{\dd}{\mathrm{d}\!}
\DeclareMathOperator{\ee}{\mathrm{e}}
\DeclareMathOperator{\ii}{\mathrm{i}}

\begin{document}

\title{Multiple scattering theory for superconducting heterostructures}

\author{G{\'{a}}bor \surname{Csire}}%
\affiliation{Institute for Solid State Physics and Optics,
             Wigner Research Centre for Physics, Hungarian Academy of Sciences, \\
             PO Box 49, H-1525 Budapest, Hungary}
\affiliation{Department of Physics of Complex Systems, E{\"{o}}tv{\"{o}}s University,
             H-1117 Budapest, P{\'a}zm{\'a}ny P{\'e}ter s{\'e}t{\'a}ny 1/A, Hungary}

\author{Bal{\'a}zs {\'U}jfalussy}
\affiliation{Institute for Solid State Physics and Optics,
             Wigner Research Centre for Physics, Hungarian Academy of Sciences, \\
             PO Box 49, H-1525 Budapest, Hungary}

\author{J{\'o}zsef Cserti}
\affiliation{Department of Physics of Complex Systems, E{\"{o}}tv{\"{o}}s University,
             H-1117 Budapest, P{\'a}zm{\'a}ny P{\'e}ter s{\'e}t{\'a}ny 1/A, Hungary}

\author{Bal{\'a}zs Gy{\H o}rffy$^\dagger$}
\affiliation{H. H. Wills Physics Laboratory, University of Bristol, Tyndall Ave, BS8-1TL, United Kingdom}

\date{\today}

\begin{abstract}

We generalize the screened Korringa-Kohn-Rostoker (SKKR) method for solving the corresponding
Kohn-Sham-Bogoliubov-de Gennes (KSBdG) equations for surfaces and interfaces.
As an application of the newly developed theory we study the quasiparticle spectrum of Au overlayers on a Nb(100) host.
We find that, within the superconducting gap region, the quasiparticle spectrum consists of
Andreev bound states (ABS) with a dispersion which is closely connected to the underlying electronic structure
of the overlayer. We also find that the spectrum has a strongly k-dependent induced gap. The properties of the gap is 
discussed in relation to the thickness of the overlayer, and it is shown that certain states do not participate in the 
Andreev scattering process. 

\end{abstract}

\pacs{31,15.A-,73.20.-r, 74.20.Pq, 74.25.Jb, 74.45.+c, 74.78.Fk}

\maketitle

\section{Introduction \label{sec:introduction}}

The theory of Bardeen, Cooper, and
Schrieffer (BCS) successfully describes the universal properties of conventional (s-wave) superconductors~\cite{BCS},
but  it can not be applied easily to inhomogeneous systems where the wave number, $k$, is not a good quantum number.
The generalization of the well-known Hartee-Fock method
with the introduction of the concept of mixed particle-hole excited states~\cite{Bogo, Valatin}
yields the Bogoliubov-de Gennes (BdG) equations~\cite{deGennes}.
In this description the standard momentum operators are replaced by field operators,
which have the advantage that they are able to describe inhomogeneous
systems.
However, this is only a mean-field theory and can not be
considered as a predictive approach to allow the computation of
material-specific properties.
For that purpose a density functional theory (DFT) was constructed by Oliveira,
Gross and Kohn (OGK)~\cite{OGK}.
In this theory the ground state energy is proved to be a
unique functional of the $\rho(\vec r)$ charge density and the
$\chi(\vec r)=\left < \Psi_\uparrow(\vec r)\Psi_\downarrow(\vec r)\right >$ anomalous density.
Later, the concept is further developed into a multicomponent density functional theory for
the combined system of electrons and nuclei (phonons)~\cite{Kreibich, Luders1, Luders2}.
The usefulness of the OGK approach~\cite{OGK} has
been demonstrated by Suvasini et al. where they introduced a simple semi-phenomenological parametrization
of the exchange-correlation functional~\cite{Suvasini}. Despite the simplicity of this approximation,
they were able to describe many features of the bulk niobium in the superconducting state, which are accessible to experiments.
Using this semi-phenomenological parametrization, one can derive a set of equations which allows the self-consistent
solution of the following coupled Kohn-Sham-Bogoliubov-de~Gennes (KSBdG) eigenvalue equations
in atomic (Rydberg) units~\cite{Suvasini}
\begin{subequations}
\begin{eqnarray}
\!\!\!\!\!\!\!\!\!\!\!\! \left(H_e(\vec{r}) - E_F \right) u_n(\vec r) +
\Delta_{eff}(\vec r) v_n(\vec r) &=&
\epsilon_n u_n (\vec r), \\
\!\!\!\!\!\!\!\!\!\!\!\! \left(H_e(\vec{r}) - E_F \right) v_n(\vec r) -
\Delta_{eff}^*(\vec r) u_n(\vec r) &=&
- \epsilon_n v_n (\vec r),
\end{eqnarray}%
\label{eq:KSBdG}%
\end{subequations}%
where $H_e(\vec{r})=-\nabla^2 + V_{eff}(\vec r)$ is the single-particle Hamiltonian,
and the wavefunction is decomposed into an electron-like part $u_n(\vec r)$
and a hole-like part $v_n(\vec r)$. The effective electrostatic and pairing potentials are
\begin{subequations}
\begin{eqnarray}
  V_{eff}(\vec r)   &=&  V_{ext}(\vec r) +
  \int \frac{\rho(\vec r')}{|\vec r - \vec r'|}  \dd^{~\!3}\!r' +
  \frac{\delta E^0_{xc}[n]}{\delta \rho(\vec r)},\\
  \Delta_{eff}(\vec r)&=& \lambda ~ \chi(\vec r),
\end{eqnarray}
\end{subequations}%
where $V_{ext}(\vec r)$ is the external potential (e.g. the Coulomb attraction from the protons).
The $\rho(\vec r)$ charge and $\chi(\vec r)$ anomalous densities can be calculated from the wavefunction components:
\begin{subequations}
\begin{eqnarray}
  \rho(\vec r) &=& 2 \sum_n \left[ |u_n(\vec r)|^2 f(\epsilon_n) +
            |v_n(\vec r)|^2 (1-f(\epsilon_n)) \right], \\
  \chi(\vec r) &=& \sum_n  u_n(\vec r)  v_n^*(\vec r) (1-2f(\epsilon_n)).
\end{eqnarray}
\end{subequations}
$f(\epsilon_n)$  is the Fermi-Dirac distribution function,
$E^0_{xc}[n]$ is the usual exchange correlation energy for normal electrons
and $\lambda$ is a semi-phenomenological adjustable parameter (it can be site-dependent).
It should be noted that the zero point of the energy scale is the Fermi level.

The past few years have shown a growing interest
in the study of superconductor based heterostructures~\cite{Yamazaki1, Yamazaki2, Ferrier}.
It is known that such inhomogeneities in the pairing potential can lead to bound quasiparticle states.
These states have been found theoretically in superconductor -- normal metal -- superconductor
heterostructures~\cite{Andreev66} and also in other systems~\cite{deGennes63,Caroli} .
The Andreev reflection~\cite{Andreev64} has been identified as the key effect which
results in such bound states, called Andreev bound states (ABS):
an electron, with energy lying in the superconducting gap, arriving from the normal metal to the
superconductor -- normal metal (S/N) interface
is retro-reflected as a hole and a Cooper pair is formed in the superconductor.
While a great many theoretical works were dedicated to study the Andreev reflection
and the ABS~\cite{Plehn,BTK,Cserti1,Cserti2,Cserti3}, it was done on model systems only,
their material specific dispersion, their "band structure" has never been calculated (nor observed experimentally) to date.
Within the framework of a tight-binding model, Annett and coworkers
also investigated such heterostructures~\cite{Annett, Fritsch}.
They have shown the existence of ABS within the gap,
and pointed out effects associated with the interplay of the gap and the normal van Hove peaks~\cite{Annett}.
The next logical step in this series of investigations is first principles calculations for real materials.
In this paper we address this problem by developing a multiple scattering theory (MST) for the solution of
the KSBdG Eqs.~(\ref{eq:KSBdG}) for surfaces
and interfaces of S/N heterostructures.
As presented in Refs.~\onlinecite{Suvasini,Plehn},
the application of constant pairing potentials gives a very good estimation
of the self-consistent solutions.
This facilitates to model a S/N system with a finite, constant
pairing potential on the superconducting host and zero on the normal metal overlayers.
In order to treat this semi-infinite geometry, a Green function based method is needed, like for example
a Screened Korringa-Kohn-Rostoker (SKKR) method~\cite{Szunyogh, Zeller} within MST.

The present paper is organized as follows. In Sec.~II we generalize the SKKR method for the solution of the
KSBdG Eqs.~(\ref{eq:KSBdG}).
Sec.~III is denoted to the computational details.
In Sec.~IV we illustrate the power of the developed method by
studying the quasiparticle "band" structure of the niobium (Nb) -- gold (Au) system.
Finally Sec.~V is devoted to the summary. Some technical details are provided in the Appendix.

\section{Generalization of the MST for superconductors: The BdG-SKKR method}

The central problem of the DFT calculations
is the solution of the KSBdG Eqs.~(\ref{eq:KSBdG}) in order to determine the single-particle
wavefunctions and the corresponding eigenvalues. However,
the single-particle Green function contains all information about the ground state.
The local DOS, the anomalous and charge densities can be obtained from the
single-particle Green function. Consequently, if the single-particle Green function is obtained,
it is not necessary to calculate the Kohn-Sham orbitals.
In this section we show how the SKKR method can be generalized
to get the single-particle Green function for multilayered systems
corresponding to the KSBdG Eqs.~(\ref{eq:KSBdG}).
Here we do not try to follow every single step of the derivation, as it would be too extensive for this paper,
 instead we just show how the most important quantities, concepts and formulas
needs to be modified due to the presence of holes. Most interim derivations can be performed in analogy of the normal state
MST, well described for example in Ref. \onlinecite{Zabloudil}.
The first step in this generalization is to decompose the BdG Hamiltonian in the following way:
\begin{equation}
 H_{BdG}(\vec r)=H_0(\vec r)+\mathcal{V}(\vec r),
 \label{eq:H_BdG}
\end{equation}
where
\begin{equation}
H_0(\vec r)=
 \begin{pmatrix}
  -\nabla^2-E_F & 0 \\
  0 & \nabla^2+E_F
 \end{pmatrix},
\end{equation}
\begin{equation}
 \mathcal{V}(\vec r)=
 \begin{pmatrix}
  V_{eff}(\vec r)  & \Delta_{eff}(\vec r) \\
  \Delta_{eff}^*(\vec r) & -V_{eff}(\vec r)
 \end{pmatrix}.
 \end{equation}
In the KKR method, the potential is usually treated in the so called muffin-tin approximation, ie.
the potential is written as a sum of single-domain potentials centered around each lattice site, $n$, namely
$V_{eff}(\vec r)= \sum_n V_n(\vec r)$.
It is usually assumed, that $V_n(\vec r)$ is spherically symmetric. This is not a necessary assumption for the theory,
however, MST for a general shape of potentials are still not common even for the normal state. Therefore, in what
follows we still restrict ourselves to spherical atomic potentials.
In our approach, we assume the same form for the effective pair interaction as well,
namely $\Delta_{eff}(\vec r)= \sum_n \Delta_{n}(r)$.
The potentials $V_n(\vec r)$  are of muffin-tin type which means that  $V_n(r)=0$
and $\Delta_n(r)=0$ if $r=|\vec r_n|\geq S_n$, where $S_n$ is the muffin-tin radius.

\subsection{Operator formalism and the free-particle Green function}

The resolvent of the BdG Hamiltonian can be defined as
\begin{equation}
 \mathcal{G}(z)=\left( z \mathbb{I}
 -H_{BdG}\right)^{-1},
 \label{eq:res}
\end{equation}
which has the usual property
\begin{equation}
\mathcal{G}\left(  z^{\ast}\right)  =\mathcal{G}\left(  z\right)  ^{\dag}\;.
\label{eq:resolvent-adjoint}%
\end{equation}
At the real axis the up- and down-side limits of $\mathcal{G}(z)$ are defined by
\begin{equation}
  \mathcal{G}(z=\epsilon\pm \ii 0)= \mathcal{G}^\pm (\epsilon).
\end{equation}
Similarly to the normal state, we can define the $\mathcal{T}$-operator as
\begin{equation}
 \mathcal{T}(z)=\mathcal{V}+\mathcal{V}\mathcal{G}(z)\mathcal{V}=\sum_{n,m} \tau^{nm}(z),
  \label{eq:T-op}
\end{equation}
where $\tau^{nm}(z)$ is the so called scattering path operator (SPO)
which comprises all possible scattering events between
the sites $n$ and $m$, including now the Andreev reflection as well.
Since $\mathcal{V}$ is Hermitean,
\begin{equation}
 \mathcal{T}(z^*)=\mathcal{T}(z)^\dag.
 \label{eq:T-res}
\end{equation}
The two different, generalized eigenfunctions of $H_{BdG}$ can be obtained from the Lippmann-Schwinger equation:
\begin{equation}
 \psi^\pm(\epsilon)= \varphi(\epsilon)+
  \mathcal{G}_0^\pm(\epsilon)\mathcal{T}^\pm(\epsilon) \varphi(\epsilon),
  \label{eq:LS}
\end{equation}
where $\varphi(\epsilon)$ is a generalized, multicomponent eigenfunction of $H_0$
and $\mathcal{G}_0(z)$ is the resolvent corresponding to $H_0$.
Here we emphasize the difference from the normal state now,
that is the wavefunctions also have a hole part.

Following the footsteps of normal state MST, we define the following orthogonal and complete basis set:
\begin{subequations}
\begin{equation}
\varphi_L^e(z,\vec r)=
  \begin{pmatrix}
    \frac{(z + E_F)^{1/4}}{\sqrt{\pi}}j_L^e(z,\vec r)\\
    0
  \end{pmatrix},
\end{equation}
\begin{equation}
  \varphi_L^h(z,\vec r)=
  \begin{pmatrix}
    0 \\
    \frac{(-z + E_F)^{1/4}}{\sqrt{\pi}}j_L^h(z,\vec r)
  \end{pmatrix},
\end{equation}
\label{eq:basis}
\end{subequations}
where $L=(l,m)$ is the usual composite index, and
\begin{equation}
 j_L^e (z,\vec r) \equiv j_{l}(p^e r) Y_L(\hat r), \quad j_L^h (z,\vec r) \equiv j_{l}(p^h r) Y_L(\hat r),
\end{equation}
$j_{l}(x)$ is the spherical Bessel function of the first type, 
\begin{equation}
  p^e=\sqrt{E_F+z}, \qquad p^h=\sqrt{E_F-z}.
\end{equation}

In a first step the free-particle Green function is derived which is related
to the structure constant describing the structural properties of the investigated system.
Using the definition of the basis abobe,  Eqs.~(\ref{eq:basis}), in terms of contour integrations
-- commonly used in MST --
it can be shown, that the Green function of free particles has the following form:
\begin{equation}
  G_0^{ab}(z,\vec r, \vec r' ) =~
  \delta_{ab}~ \sum_L ~
  H_{L}^a(z,\vec r_>)\left[{j}^a_L (z,\vec r_<)  \right]^\times
\label{eq:G0}
\end{equation}
where
\begin{equation}
 H_{L}^a(z,\vec r_n) \equiv -\ii p^a  h_{L}^a(z,\vec r_n)
 \equiv -\ii p^a  h_{l}^a(p^a r_n) Y_L(\hat r_n).
 \label{eq:def_H}
\end{equation}
Here we used the notation $r_{<}=\min\left(  r,r^{\prime}\right)  $,
$r_{>}=\max\left(  r,r^{\prime}\right) $ with 
the choice of Im$\left\{p^e\right\}>0$ and Im$\left\{p^h\right\}>0$,
$a,b$ denote the electron, hole indices, and 
$h_{l}^e(x)=h_{l}^+(x)$ is the Hankel function of the first type, while
$h_{l}^h(x)=-h_{l}^+(x)$. 
We defined the $\times$ operator for any arbitrary function, $f$, as
$f_L^a (z,\vec r_n)^\times \equiv f_{l}(p^a r_n) Y_L(\hat r_n)^*$.
Therefore, it is clear, that $G_0^{ab}$ is diagonal in indeces $a$ and $b$, and 
the hole part of the free-particle Green function can be obtained
from its electronic part
\begin{equation}
G_{0}^{hh}\left(  z;\vec{r},\vec{r}^{\prime}\right) = -
G_{0}^{ee}\left( - z;\vec{r},\vec{r}^{\prime}\right).
\label{eq:G0hh}
\end{equation}

\subsection{Scalar relativistic Bogoliubov-de Gennes equations}

Nowadays almost all electronic structure codes are built around what is called the "scalar relativistic" approximation,
where the mass-velocity and Darwin terms are properly taken into consideration, but the spin-orbit
coupling is neglected.
Consequently, to be able to thoroughly compare our results with normal state electronic structure calculations,
a scalar relativistic generalization of the BdG theory is needed.
To arrive to such a theory, one needs to start from the
relativistic Dirac-Bogoliubov-de Gennes (DBdG) equations already worked out in the literature~\cite{Capelle1, Capelle2}.
An analogous scalar relativistic form of the BdG equations can be obtained quite straightforwardly by neglecting not only
the spin-orbit coupling term but all relativistic corrections to the pairing potential as well. 
By suppressing the explicit dependence on the complex energy, on a log-scale ($x=\log r$) these coupled equations for the radial part
-- since both the potential and the pairing potential is spherically symmetric -- can be written as follows:
\begin{subequations}
\begin{align}
  \frac{\dd}{\dd x} Q_l^e(x) & = -Q_l^e(x) + U_l^e(x) P_l^e(x) + \ee^x \Delta(x) P_l^h(x), \\
  \frac{\dd}{\dd x} P_l^e(x) & = P_l^e(x) + \ee^x B^e(x) Q_l^e(x), \\
  \frac{\dd}{\dd x} Q_l^h(x) & = -Q_l^h(x) + U_l^h(x) P^h(x) - \ee^x \Delta^* (x) P_l^e(x), \\
  \frac{\dd}{\dd x} P_l^h(x) & = P_l^h(x) + \ee^x B^h(x) Q_l^h(x) ,
\end{align}
\label{eq:SRBdG}
\end{subequations}
where the wavefunctions are defined as
\begin{equation}
  \begin{pmatrix}
  P_l^e(x) \\
  P_l^h(x)
\end{pmatrix}= \ee^x \begin{pmatrix}
  u_l(x) \\
  v_l(x)
\end{pmatrix},
\end{equation}
and
\begin{subequations}
\begin{align}
  U_l^e(x) & = \frac{l(l+1)}{\ee^x B^e(x)} + \ee^x (V(x)-z), \\
  U_l^h(x) & = \frac{l(l+1)}{\ee^x B^h(x)} + \ee^x (V(x)+z),\\
  B^e(x) & = 1+ \frac{z-V(x)}{c^2}, \\
  B^h(x) & = 1- \frac{z +V(x)}{c^2}.
\end{align}
\end{subequations}
In MST, the scattering matrices and the scattering solutions are obtained by matching the solutions
of the above equations inside the muffin-tin sphere to the solutions outside.

\subsection{Single-site scattering}

After performing the necessary integrations in the Lippmann-Schwinger equation (\ref{eq:LS})
and using the basis defined in Eqs.~(\ref{eq:basis}) together with the definition of the
free-particle Green function Eq.~(\ref{eq:G0}), two different scattering solutions can be obtained and written
in the following matrix-form
\begin{equation}
 R_L^{n,ab}(z, \vec r_n)= j_L^a (z,\vec r_n) \delta_{ab}
+ \sum_{L'} H_{L'}^b(z,\vec r_n) t^{n,ba}_{L'L}(z),
\label{eq:sol}
\end{equation}
where ${t}^{n,ab}_{L'L}(z)$ is the single-site t-matrix
which is diagonal in $L,L'$ indices for potentials with spherical symmetry.
Equation~(\ref{eq:sol}) implies that an electron (hole) like solution to
the Lippmann-Schwinger equation may have a hole (electron) like component as well.
If the incoming wave is electron like, the solution outside the muffin-tin sphere can be written as
\begin{subequations}
\begin{equation}
 R_L^{e} (z,\vec r) =
 \begin{pmatrix}
  j_l(p^e r) - \ii p^{e\,} t_{l}^{ee} (z) h_l^+ (p^e r) \\   \ii p^{h\,} t_{l}^{he}(z) h_l^+ (p^h r)
 \end{pmatrix}
Y_{lm}(\theta, \phi)
\end{equation}
and if the incoming wave is hole like, it can be expressed as
\begin{equation}
 R_L^{h} (z,\vec r) =
 \begin{pmatrix}
  - \ii p^{e\,} t_{l}^{eh}(z) h_l^+ (p^e r) \\
   j_l(p^h r) + \ii p^{h\,} t_{l}^{hh}(z) h_l^+ (p^h r)
 \end{pmatrix}
Y_{lm}(\theta, \phi).
\end{equation}
\end{subequations}
It should be noted that these equations are different not only in the electron-hole components, but also  in the appropriate
energy dependence as well through $p^e$ and $p^h$.
As mentioned earlier, the t-matrix can be obtained by matching the outside scattering solutions and the regular solutions inside
the muffin-tin sphere at the boundary, which is described in more details in the Appendix.

Using the particle-hole symmetry~\cite{deGennes}, it can be easily proved that
\begin{subequations}
\begin{eqnarray}
 t_l^{he}(-z) &=& t_l^{eh}(z),
 \label{eq:the} \\
 t_l^{ee}(-z) &=& -t_l^{hh}(z).
 \label{eq:tee}
\end{eqnarray}
\label{eq:tsym}
\end{subequations}
These symmetry relations are independent from the actual form of $\Delta(r)$ and
$V(r)$ in the superconducting muffin-tin sphere.

\subsection{Multi-site scattering, generalized Faulkner-Stocks formula}

A rather convenient property of the KKR method that it allows a transparent decoupling of the potential (described by its scattering matrix) 
and the structural properties of the system of scattering centers (atoms).
Similarly to the normal case~\cite{Zabloudil},
using the well-known expansion of plane waves into spherical Bessel function and spherical harmonics
(Bauer's identity~\cite{Zabloudil}) and the definition of the free-particle Green function Eq.~(\ref{eq:G0}),
the free, real space structure constants to describe the structural properties can be constructed as:
\begin{equation}
  G_{0,LL'}^{nm,ab}(z) =  \delta_{ab} 4 \pi \sum_{L''} \ii^{L-L'-L''}
  H_{L''}^a(z,\vec R_{nm}) C_{LL''}^{L'},
  \label{eq:struc_const}
\end{equation}
where $\vec R_{nm}$ is the vector pointing from site $n$ to site $m$ and $C_{LL''}^{L'}$ are the usual Gaunt-coefficients~\cite{Zabloudil}.

The scattering matrices, the matrices of the structure constant and the scattering path operator can be introduced in 
a quasiparticle-site-angular momentum representation,
\begin{eqnarray}
 \mathbf{t}(z) &=& \{t^{n,ab}_{LL'}(z) \delta_{nm} \},
 \label{eq:sh_t}\\
 \mathbf{G}_0(z) &=& \{G^{nm,ab}_{0,LL'}(z) (1-\delta_{nm})\},
 \label{eq:sh_G_0}\\
 \boldsymbol{\tau}(z) &=& \{\tau^{nm,ab}_{LL'}(z) \},
 \label{eq:sh_tau}
\end{eqnarray}
where the $\boldsymbol{\tau}(z)$ SPO can be determined from the single-site t-matrix
and the real space structure constant similarly to normal state in the supermatrix formalism
\begin{equation}
 \boldsymbol{\tau}(z)=
 \left(
 \mathbf{t}(z)^{-1} -\mathbf{G}_0(z)
 \right)^{-1}.
  \label{eq:sh_tau_t}
\end{equation}
For periodic systems  in KKR theory it is possible to write the above equations in reciprocal space, which allows to obtain
the SPO as a function of the wave number and the energy. Finding the poles of the SPO
as a function of $k$ and $\epsilon$ gives the electronic band structure.
Butler described a one-dimensional version of KKR~\cite{Butler}, which is
often used as a testbed for new ideas within the theory. This has been done in Ref.~\onlinecite{Suvasini92}, where
a one-dimensional model of the Bogoliubov-de Gennes-KKR theory has been presented.
However, since translational invariance is broken at the interface, to be able to calculate physical
properties on surfaces and interfaces, we follow the derivation of a full real-space Green function,
and make use of two-dimensional periodicity later.

In the normal state MST, it had been shown by Faulkner and Stocks~\cite{Faulkner} that the Green function can be obtained
from the scattering
path operator and from the scattering solutions. The derivation can be followed step by step,
and the result for the site-diagonal part of the Green function can be written in an
analogous form as well.
First, the full Green function was evaluated for the case of
$|\vec{r}_{n}|>S_{n}$, $|\vec{r}_{m}|>S_{m}$:
\begin{equation}
 G(z,\vec r, \vec r)=
   \mathbf{Z}(z,\vec r) \boldsymbol{\tau}(z) \mathbf{Z}(z,\vec r)^\times
 - \mathbf{Z}(z,\vec r) \mathbf{J}(z,\vec r)^\times,
\label{eq:genFS}
\end{equation}
where we used the following matrix notation
($\mathbf{F}=\mathbf{Z},\textrm{or }\mathbf{J}$):
\begin{subequations}
\begin{equation}
\begin{split}
& \mathbf{F}\left(z;\vec{r}\right)  \equiv
\left\{ \mathsf{F}^{n,ab}\left(z;\vec{r}\right) \right\}  \\ & \equiv
\left\{ \left[
\begin{array}
[c]{cccc}%
f_{1}^{n,ab}\left(  z;\vec{r}\right)  , & f_{2}^{n,ab}\left(  z;\vec{r}\right)  , &
f_{3}^{n,ab}\left(  z;\vec{r}\right)  , & \cdots
\end{array}
\right] \right\}  \;,
\end{split}
\end{equation}
and the corresponding adjoint vector,%
\begin{equation}
\mathbf{F}\left(z;\vec{r}\right)^\times \equiv
\left\{ \widetilde{\mathsf{F}}^{n,ab}\left(z;\vec{r}\right)^\times \right\}  \equiv
\left\{ \left[
\begin{array}
[c]{c}%
\widetilde{f}_{1}^{n,ab}\left(   z;\vec{r}\right)  ^{\times}\\
\widetilde{f}_{2}^{n,ab}\left(  z;\vec{r}\right)  ^{\times}\\
\widetilde{f}_{3}^{n,ab}\left(   z;\vec{r}\right)  ^{\times}\\
\vdots
\end{array}
\right]\right\}  \;,
\end{equation}
\end{subequations}
where
\begin{subequations}
\begin{eqnarray}
    J_l^{ab}(z,r) &=& j_l^a (z,r) \delta_{ab},\\
    Z_l^{ab}(z,r)&=& \sum_{c}  R_l^{ac}(z,r)\left[t^{-1}_l\right]^{cb}\!\!(z),
    \\
    \widetilde{Z}_l^{ab}(z,r)&=& \sum_{c} \left[t^{-1}_l\right]^{ac}\!\!(z) \widetilde{R}_l^{cb}(z,r),
    \\
    R_l^{ab}(z,r)&=& J_l^{ab} (z,r) + H_{l}^{a}(z,r) t^{ab}_{l}(z),\\
    \widetilde{R}_l^{ab}(z,r)&=& J_l^{ab} (z,r)
      + t^{ab}_{l}(z) H_{l}^{b}(z,r).
  \end{eqnarray}
  \label{eq:GF_calc}
\end{subequations}

To calculate physical quantities, we have to continue the Green function inside the muffin-tin spheres by using the solutions
of the scalar relativistic BdG Eqs.~(\ref{eq:SRBdG}) as described in details in the Appendix.
The formulas given above can be applied to surfaces and interfaces 
quite straightforwardly following the idea of the so called Screened KKR (SKKR)
formalism described in Refs.~\onlinecite{Szunyogh, Zeller}.
In this formalism, a special reference system is used to obtain
structure constants that are localized in real space.
In the supermatrix formalism we used above, the screening transformation can be written in a way that is
formally exactly the same as it was presented in Ref.~\onlinecite{Zeller}.
Thus the whole formalism can be derived for layered systems with two-dimensional
periodicity and applied as the SKKR method prescribes.

To perform fully self-consistent calculations for S/N systems,
it is necessary to calculate the  charge density and the anomalous density for layer $I$,
which can be obtained from the Green function:
\begin{widetext}
\begin{subequations}
\begin{eqnarray}
  \rho_I(\vec r) &=&
      -\frac{1}{\pi} \int_{-\infty}^{\infty} \dd  \epsilon f(\epsilon) \int_{\mathrm{BZ}} \dd^2 k_{||}
       ~\textrm{Im}~ \mathrm{Tr}_{a,L}~ G^{ab,II+}_{LL'}(\epsilon, \vec r,\vec{k}_{||}), \\
    \chi_I(\vec r) &=&
      -\frac{1}{2\pi} \int_{-\infty}^{\infty} \dd  \epsilon (1-2f(\epsilon))\int_{\mathrm{BZ}} \dd^2 k_{||}
       ~\textrm{Im}~\mathrm{Tr}_{L}~ G^{he,II,+}_{LL'}(\epsilon, \vec r, \vec{k}_{||}). 
\end{eqnarray}
\end{subequations}
\end{widetext}

\section{Computational details}

In this section, we describe the technical details of the calculation of quasiparticle spectrum
for a real superconducting heterostructure using the BdG-SKKR method outlined in Sec. II.

The geometry of our system builds up from two-dimensional translational invariant layers.
The system comprises three regions: (i) semi-infinite bulk (Nb); (ii) the interface region
that -- in our case -- consists of six superconducting layers (Nb), various number of normal metal layers (Au) and
three layers of empty spheres; (iii) and semi-infinite vacuum.
The Nb has the body-centered cubic (BCC) crystal structure with a lattice parameter $a = 3.3$~\AA.
Here we do not try to investigate the effect of matching different lattice structures on the quasiparticle spectrum.
Thus, for simplicity we assume BCC epitaxial growth for the Au overlayers and
the Nb/Au BCC(100) heterostructure will be investigated.

As we mentioned in the introduction, we do not calculate self-consistently the $\Delta_{eff}(r)$
pairing-potential, only the normal state calculation is performed self-consistently to obtain the $V_{eff}(r)$
effective potential. We do this to simplify our first calculations, and because it has been shown in Ref.~\onlinecite{Suvasini}, 
that a good guess of the self-consistent pairing-potential can be the following average
\begin{equation}
  \overline{\Delta} =\frac{1}{V_{WS}}\int_{V_{WS}} \Delta_{eff} (r) \dd r,
\end{equation}
where $V_{WS}$ is the volume of Wigner-Seitz cell.
Consequently, we treat the $\overline{\Delta}$ averaged pairing potential as an adjustable parameter.
Since $\overline{\Delta}$ is the experimentally observed gap (the gap is measured from the Fermi level), in principle, it
should be set to equal the experimental value~\cite{Pronin}. However,
with this value of the $\overline{\Delta}$ (orders of magnitude smaller (meV) than the
electronic energies (eV) involved in a normal state band structure calculation),
many layers are necessary to see its effect on the bands crossing (not just near the Fermi level), which
significantly increases the computational time.
Therefore, a model $\overline{\Delta}$ is used here to explore the quasiparticle spectrum.
The conclusions we draw, however, does not depend on the size of the $\overline{\Delta}$ parameter.

Similarly to normal state electronic structure calculations, single-site t-matrices are obtained for each layer,
where the $\overline{\Delta}$ averaged pair interactions can be different on each layer, just as the atomic potentials.
In our model a finite, constant $\overline{\Delta}$ pairing potential is assumed on the Nb layers
and $\overline{\Delta}=0$~Ry on the Au overlayers.

In practice, we obtain the t-matrix and the wavefunctions in the following way: first, the
radial scalar relativistic BdG Eqs.~(\ref{eq:SRBdG}) are integrated outwards up to
the radius of the muffin-tin sphere with different starting values to obtain the $R_l^{ab}(\epsilon,r)$.
The matching to the scattering solutions (details in the Appendix) yields the t-matrix.
Then the $H_l^{ab}(\epsilon,r)$ irregular wavefunction is calculated similarly by an integration inwards
starting at the muffin-tin radius.
The integrations are performed with a predictor-corrector algorithm~\cite{Zabloudil}
on logarithmic scale with 721 radial mesh points in the muffin-tin sphere.
To obtain the normal self-consistent potential, $V_{eff}(r)$, the energy integrals
are performed by sampling 16 points on a semicircle contour in the upper complex energy plane.
The calculations are carried out within the atomic sphere approximation with an
angular momentum cutoff of $l_{max} = 2$.
We use 2450 $k$ points for integration over the Brillouin zone to calculate the DOS of bulk Nb.

In what follows, we calculate the DOS and the Bloch spectral function (BSF) which is equivalent to the
quasiparticle spectrum. In all of the following plots the energy is measured in units of Rydbergs and $k$ in units of $\pi/a$.
The contour plots of the spectral functions are calculated in 400 energy points $\times$ 265 $k$-points.

\section{Results}

The BSF is defined as $A_B(\epsilon, k)= \sum_n \delta(\epsilon -\epsilon_n (k))$
and can be calculated directly from the Green function. In a layered system for layer $I$, this can be
expressed as
\begin{equation}
  A_B^I(\epsilon, k_{||})=
    -\frac{1}{\pi}
    \mathrm{Im}~\mathrm{Tr}~ G^{+}_{II}(\epsilon, \vec r,k_{||}).
\label{eq:BSF}
\end{equation}
Since the BSF is equivalent to the quasiparticle spectrum, drawing a contour plot of
the BSF as a function of energy along specified directions of $k$ is a powerful tool to visualize the quasiparticle states.
In a layered system this can be done for each layer, based on Eq.~(\ref{eq:BSF}).
The spectral functions were calculated by adding a small imaginary part of 0.0005~Ry to the energy.

\subsection{DOS of bulk Nb}

\begin{figure}[hbt!]
   \includegraphics[width=0.5\linewidth]{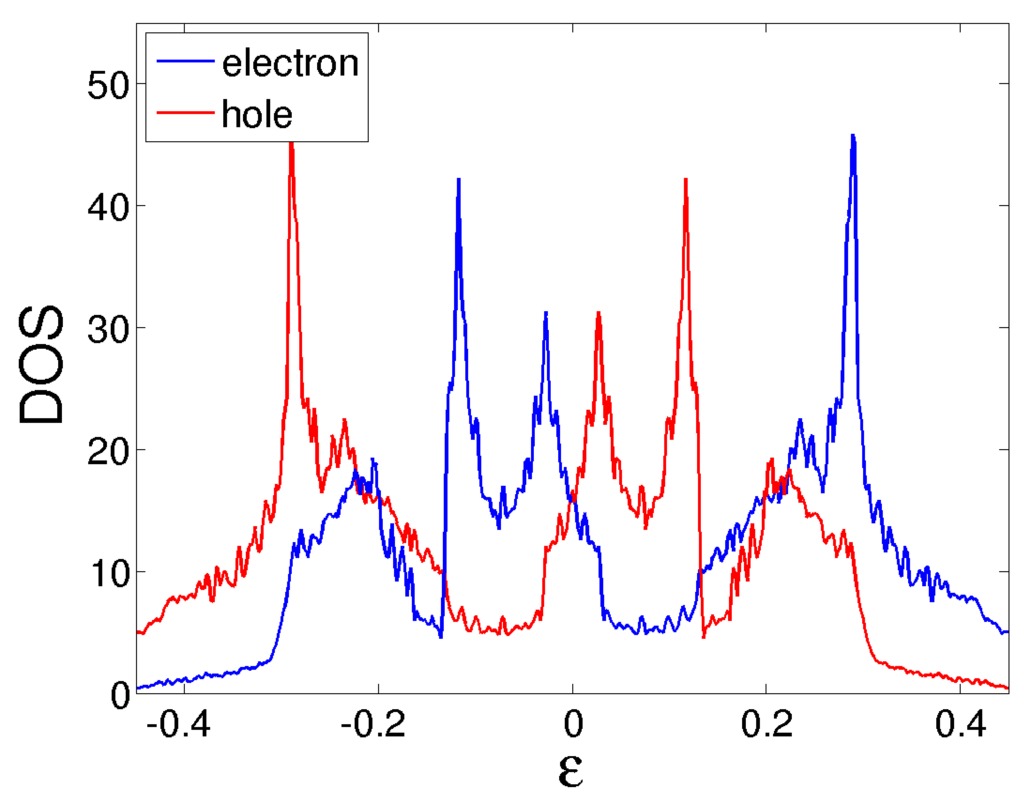}~
   \includegraphics[width=0.5\linewidth]{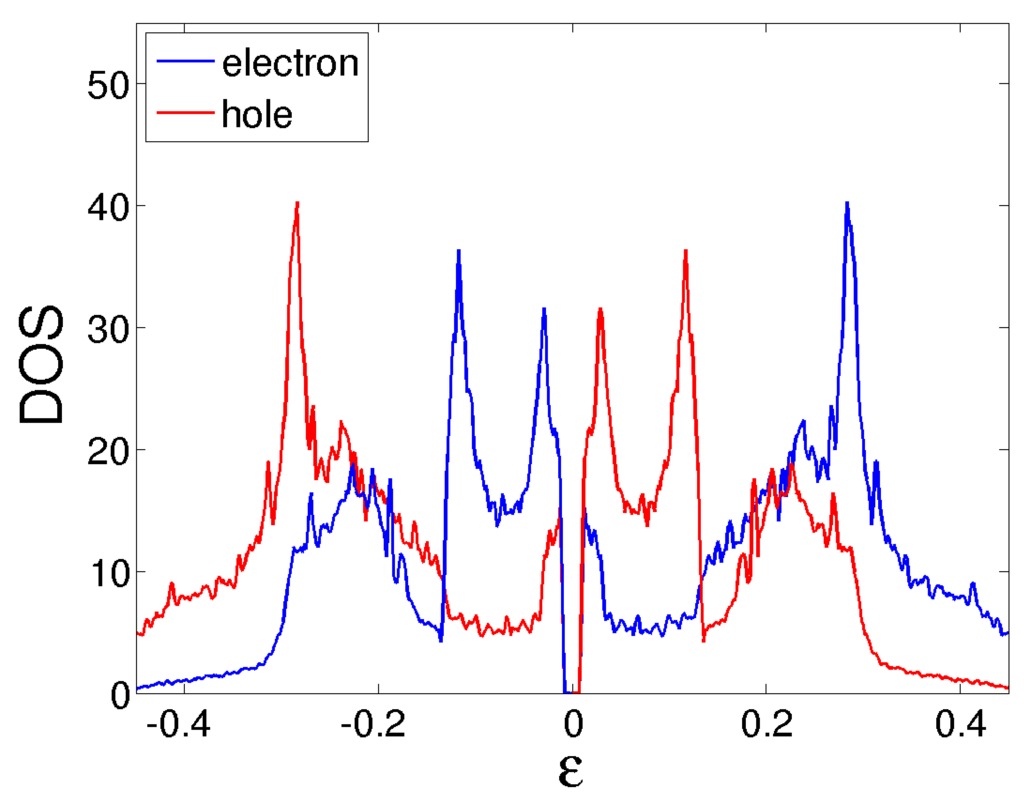}
   \caption{\label{fig:dos}%
            (Color online) DOS (arbitrary units) of bulk Nb ($E_F=0.713$~Ry)
            in the case of $\overline{\Delta}_{Nb}=0$~Ry (left panel) and $\overline{\Delta}_{Nb}=0.01$~Ry (right panel).
            The blue line corresponds to the density of electron-like states and the red to
            the density of hole-like states.}
\end{figure}

To test our procedures, and
to show the effect of the $\overline{\Delta}$ pairing potential on a bulk sytem (Nb), we first
performed calculations for the case of bulk niobium using the values $\overline{\Delta}_{Nb}=0$~Ry and
$\overline{\Delta}_{Nb}=0.01$~Ry.
The DOS can be calculated from the BSF  $D(\epsilon)=\int A_B(\epsilon, k_{||}) \dd k_{||}$.
The particle-hole symmetry implies that the density of the hole-like
states are just the reflection of the density of electron-like states to the Fermi energy. This is indeed the case in our
calculations, as it can be seen in Fig.~\ref{fig:dos} for the case of $\overline{\Delta}_{Nb}=0$~Ry (left panel).
If $\overline{\Delta}_{Nb}$ is nonzero, a gap appears around the Fermi level and the size of the gap equals to the value of
$\overline{\Delta}_{Nb}$.

\subsection{Normal state band structure of Nb/Au heterostructures}

\begin{figure}[hbt!]
   \includegraphics[width=0.52\linewidth]{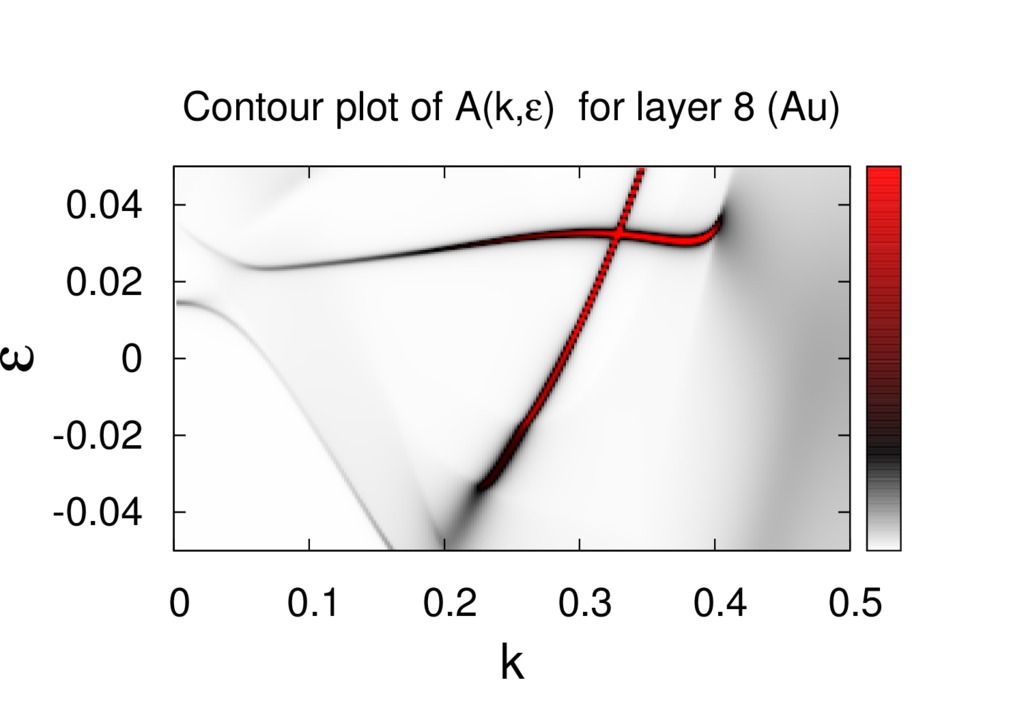}~
   \includegraphics[width=0.52\linewidth]{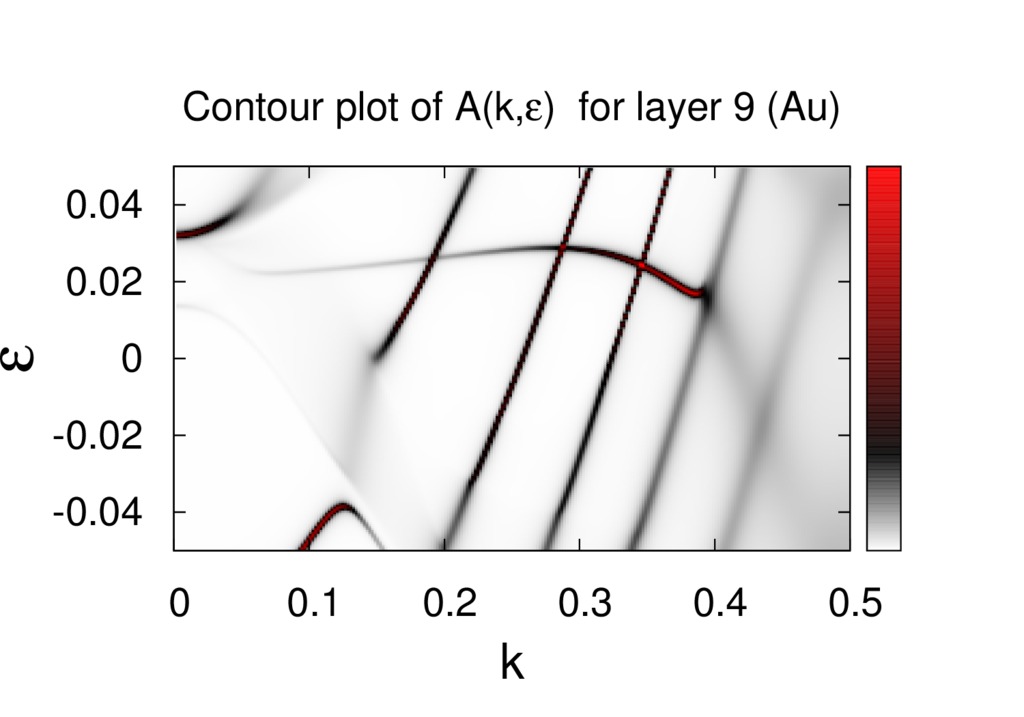}\\
   \includegraphics[width=0.52\linewidth]{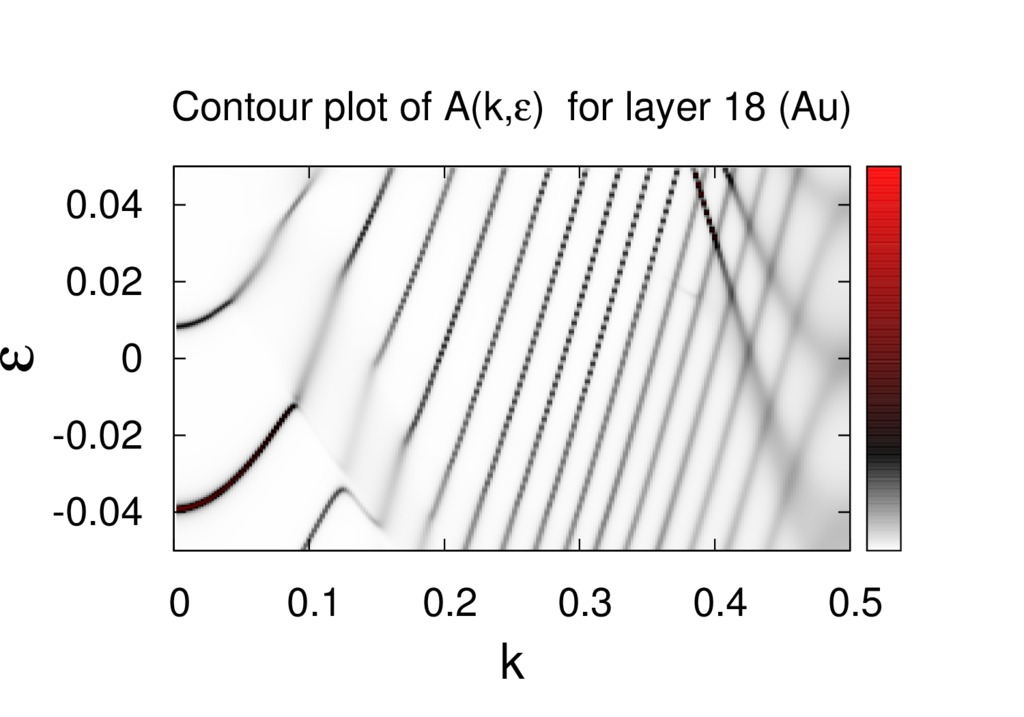}~
   \includegraphics[width=0.52\linewidth]{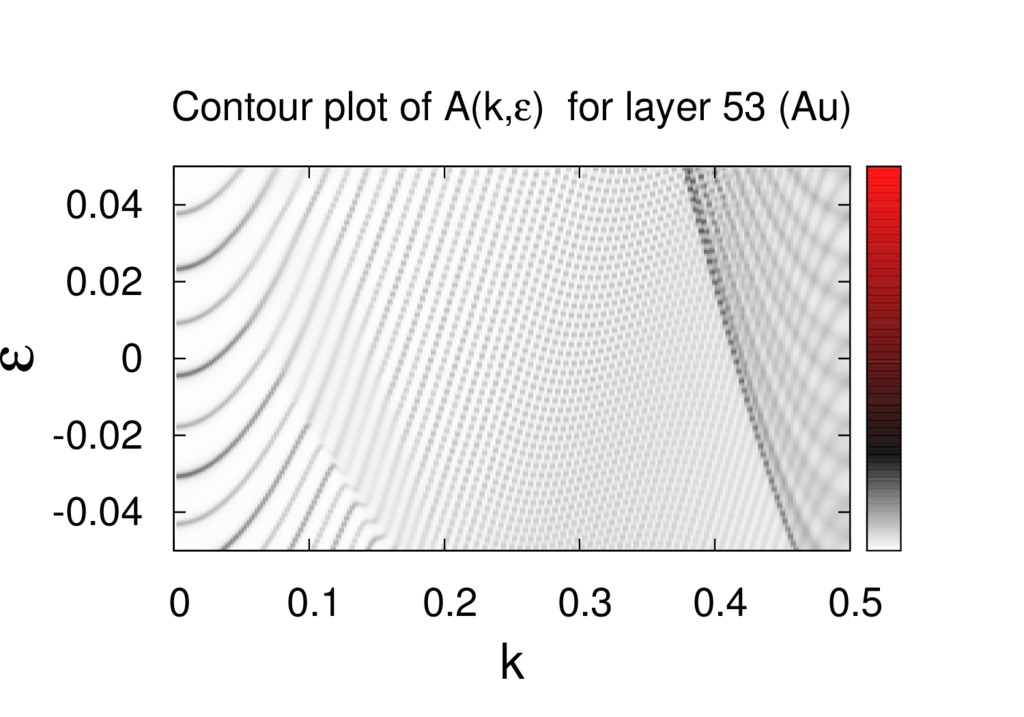}
   \caption{\label{fig:normal1}%
            (Color online)
            Contour plot of the BSF (normal state band structure)
            from the "middle" of the Au layers for different thicknesses of the Au:
            3 Au layers (top left panel) and 9 Au layers (top right panel)
            24 Au layers (bottom left panel) and 93 Au layers (bottom right panel).}
\end{figure}

\begin{figure}[hbt!]
   \includegraphics[width=0.52\linewidth]{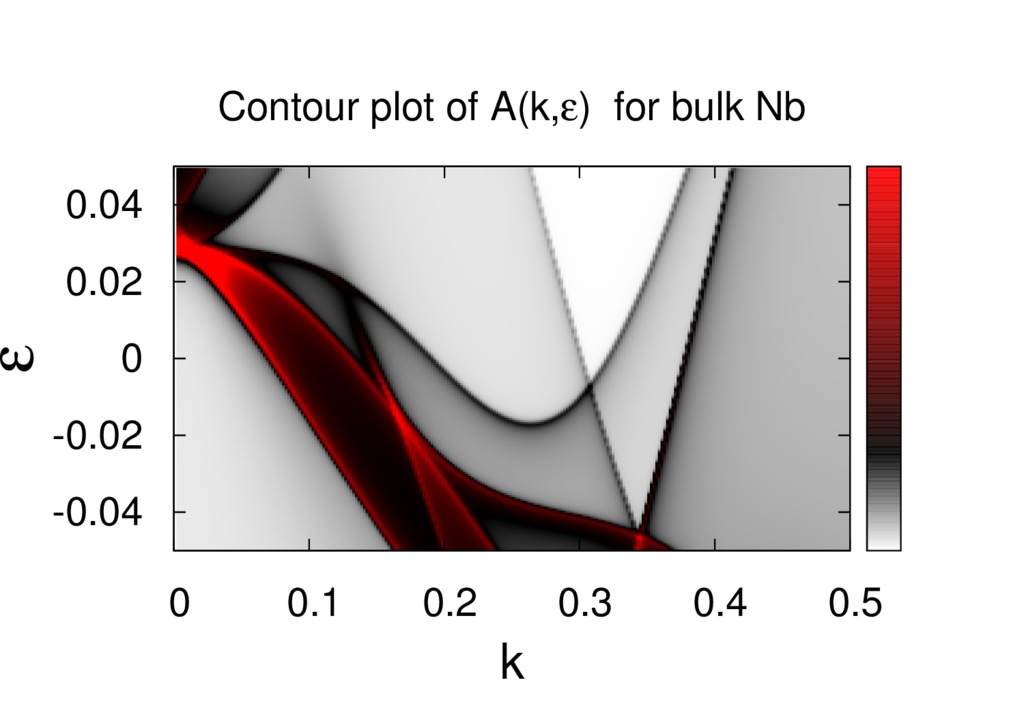}~
   \includegraphics[width=0.52\linewidth]{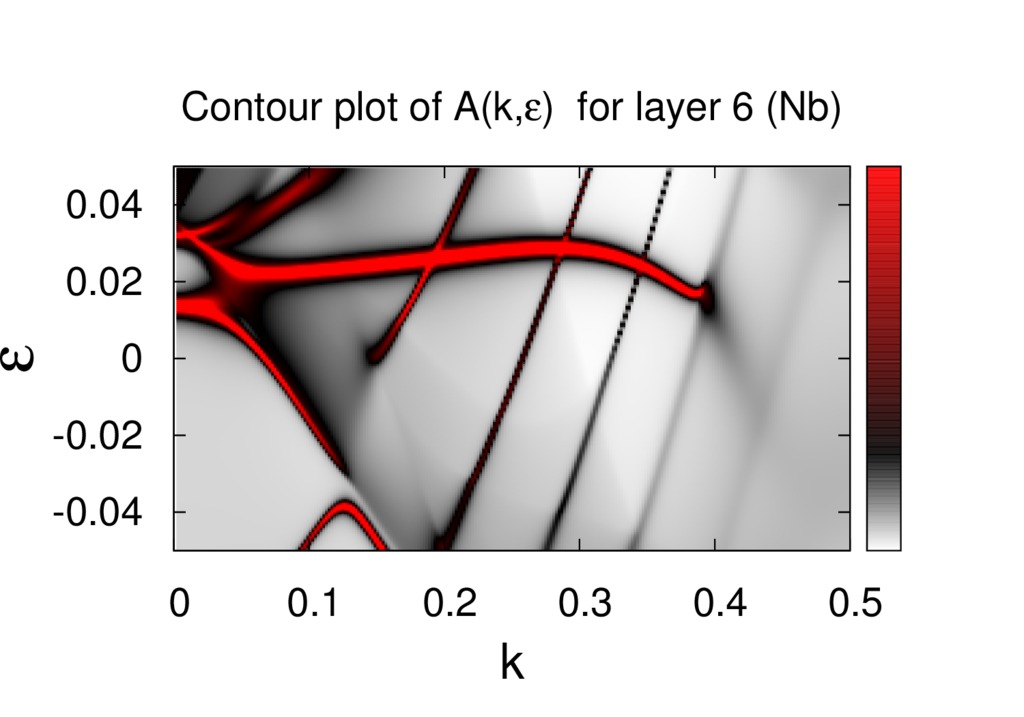}\\
   \includegraphics[width=0.52\linewidth]{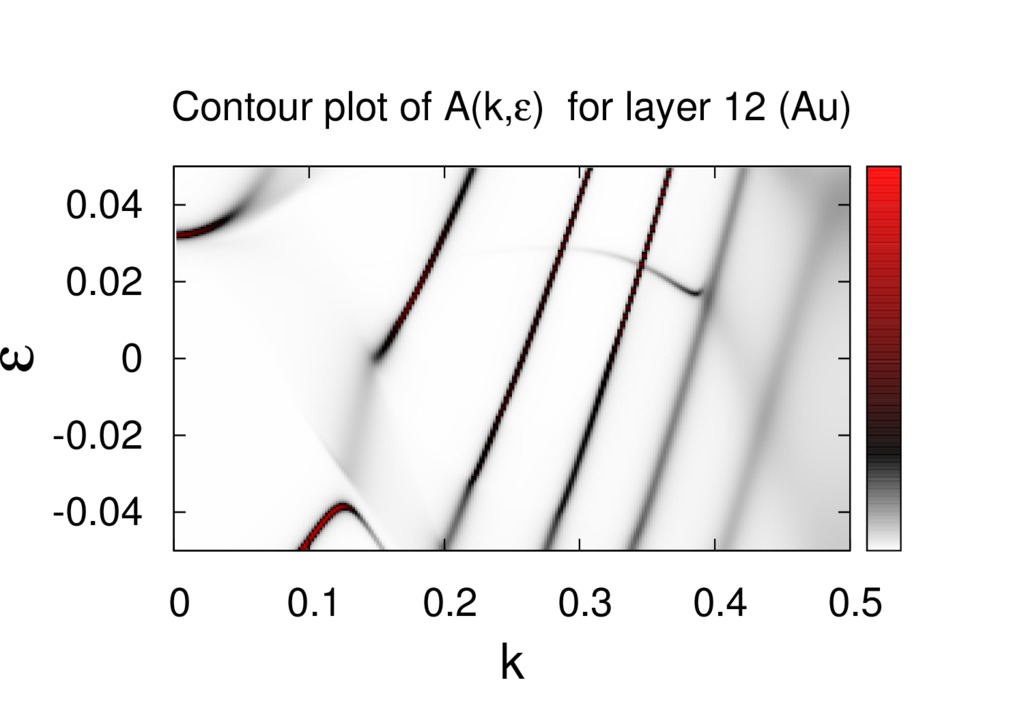}~
   \includegraphics[width=0.52\linewidth]{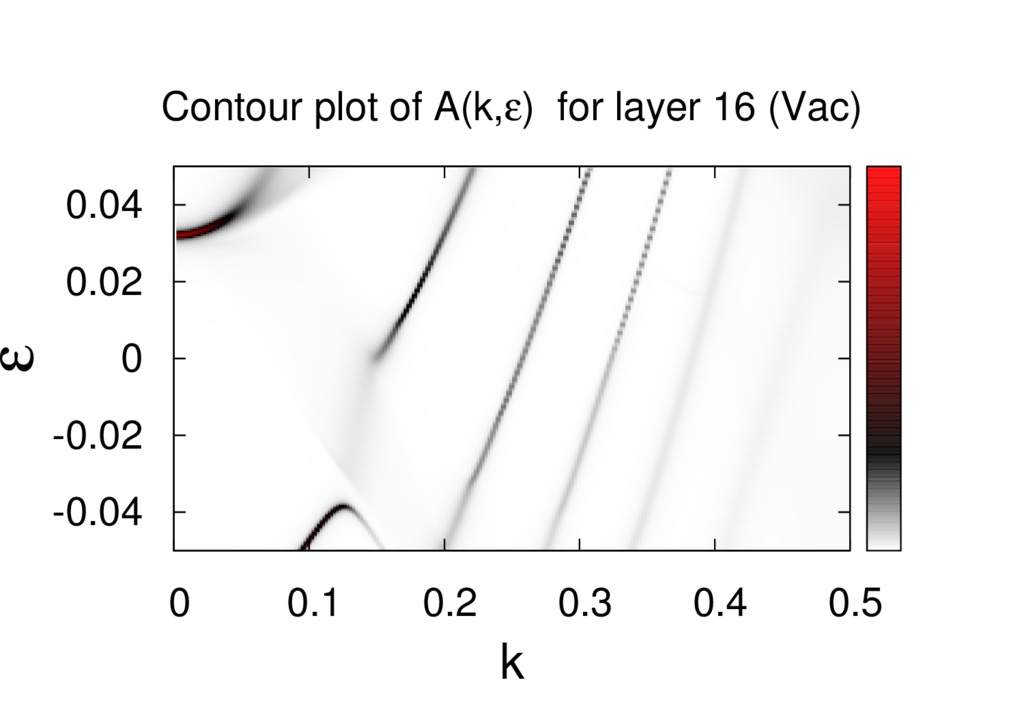}
   \caption{\label{fig:normal2}%
            (Color online)
            Contour plot of the BSF (normal state band structure) for different layers.
            In the interface region there are 6 Nb layers, 9 Au layers, 3 vacuum layers. }
\end{figure}

To demonstrate the power of our new theory for an inhomogeneous system, we apply it to study the system of Au overlayers on Nb(100).
Foremost, we made calculations for the normal state, for two reasons.
First, to obtain self-consistent potentials and work functions for the BdG calculations.
Second, it is important to explore the features in the normal state electronic structure to later understand the
quasiparticle spectrum we are planning to calculate.
Therefore, self-consistent calculations were performed for systems containing a semi-infinite Nb, an additional 6 Nb layers 
and subsequently 3, 9, 24, and 93 Au layers.
In Fig.~\ref{fig:normal1} we show the contour plot of the BSF
for a layer that we considered to be in the "middle" of the appropriate sample and for a layer in the bulk Nb
(seen in Fig.~\ref{fig:normal2} top left panel). It should be mentioned that the later is just the projection of the bulk 
spectral function on the
(100) plane and it represents the corresponding projection of the bulk band structure.
The plots are restricted in energy to the range of [-0.05~Ry,0.05~Ry] (later we will choose
$\overline{\Delta}_{Nb}$ to equal this value, and solve the BdG equations within this energy range).
When Fig.~\ref{fig:normal1} viewed as a sequence, one can immediately recognize the signatures of
confinement. Where the DOS in the bulk Nb is low, the states in the Au are confined, as they can not scatter into the Nb,
and on the other side the system is limited by vacuum. In regions where the DOS is high in the Nb, the states in the Au
are smeared out, as here the appropriate electrons can scatter more easily into the other side of the interface. The confined states
in the Au, therefore, can be regarded as quantum-well (QW) states. The confinement causes a roughly
$2\pi/L$ sampling  (where $L$ is the thickness of the Au sample) of the Au band structure.
It can be seen from the figure, that as
$L$ increases, the QW bands become denser and denser. As $L$ approaches infinity, the bulk electronic structure of
Au is recovered in the middle of the sample.

For a fixed number of Au layers, one can investigate the layer dependence of the electronic structure.
This is illustrated for the system with 9 Au layers in Fig.~\ref{fig:normal2}.
First, we have to notice that the QW bands do extend into the self-consistent Nb layers, as these layers show signatures of both
of the bulk Nb and the confined states of Au. Surprisingly, around the actual interface there is a very sharp horizontal band,
which can
be seen only at the interface layers and quickly disappears further away. It is not present either in the bulk Nb,
nor in the Au electronic structure, therefore, it may be regarded as an interface state.

\subsection{The quasiparticle spectrum of Nb/Au heterostructures}

\begin{figure}[hbt!]
   \includegraphics[width=0.52\linewidth]{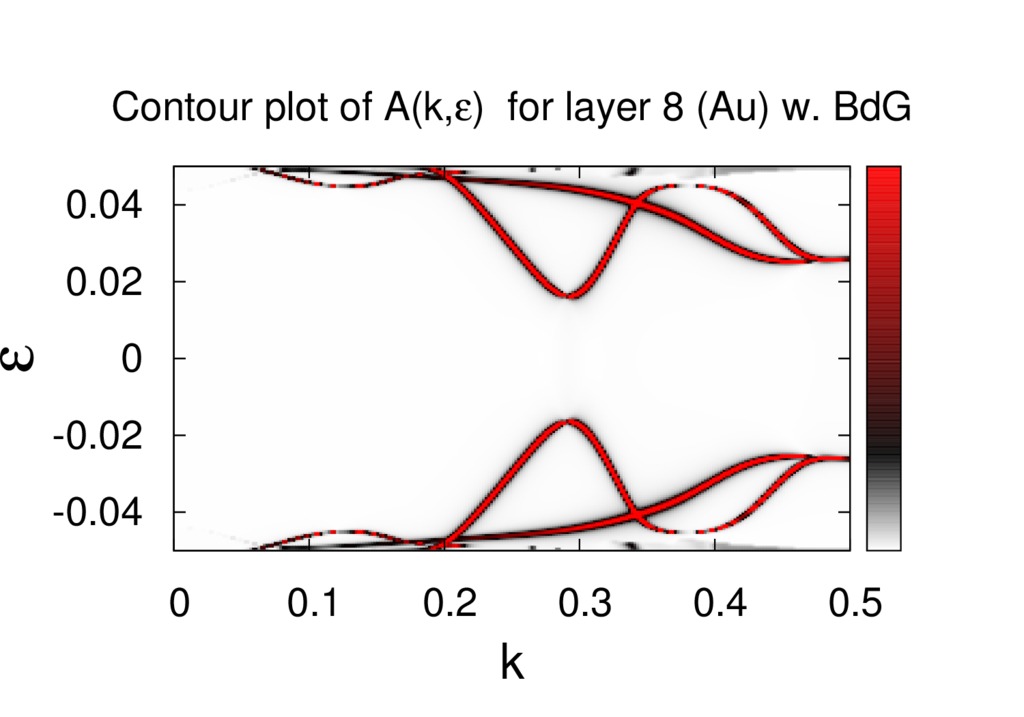}~
   \includegraphics[width=0.52\linewidth]{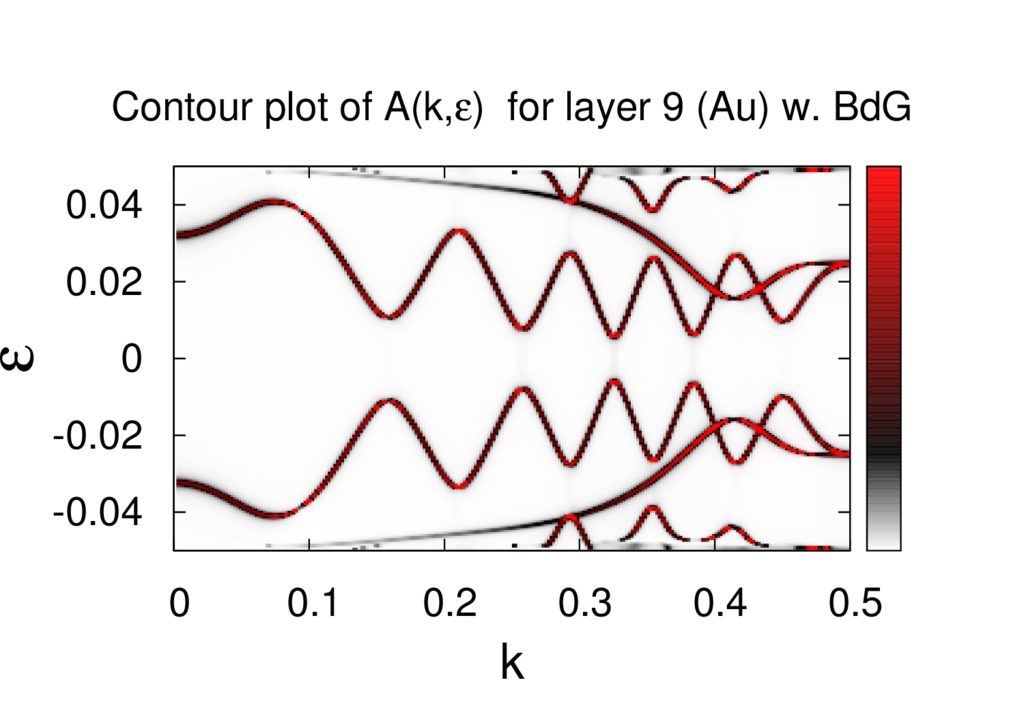}\\
   \includegraphics[width=0.52\linewidth]{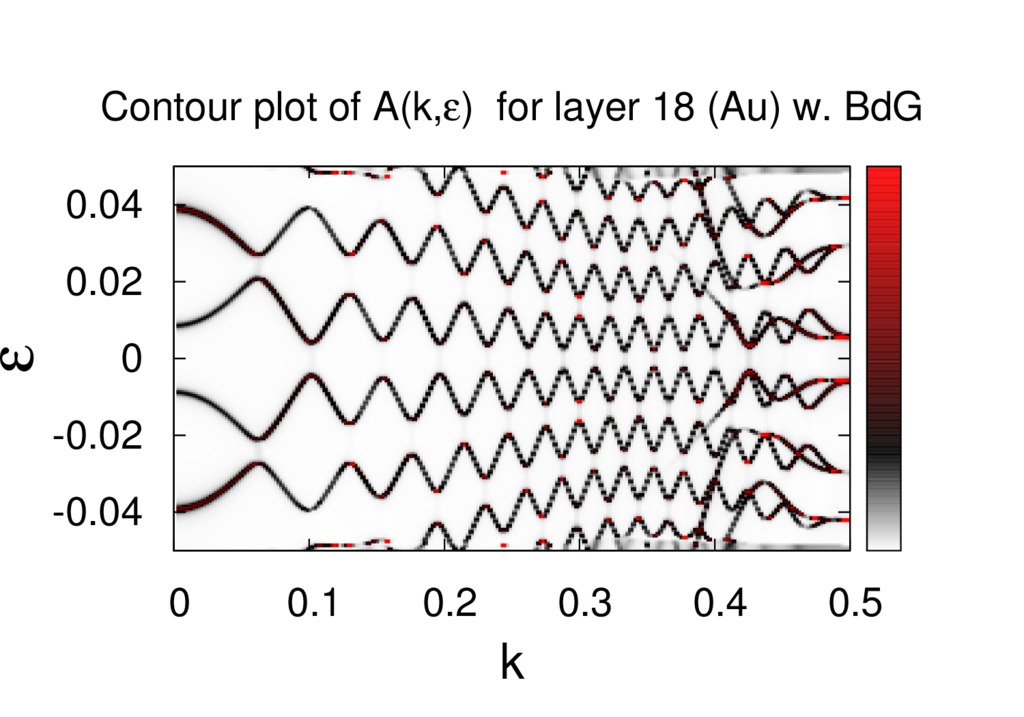}~
   \includegraphics[width=0.52\linewidth]{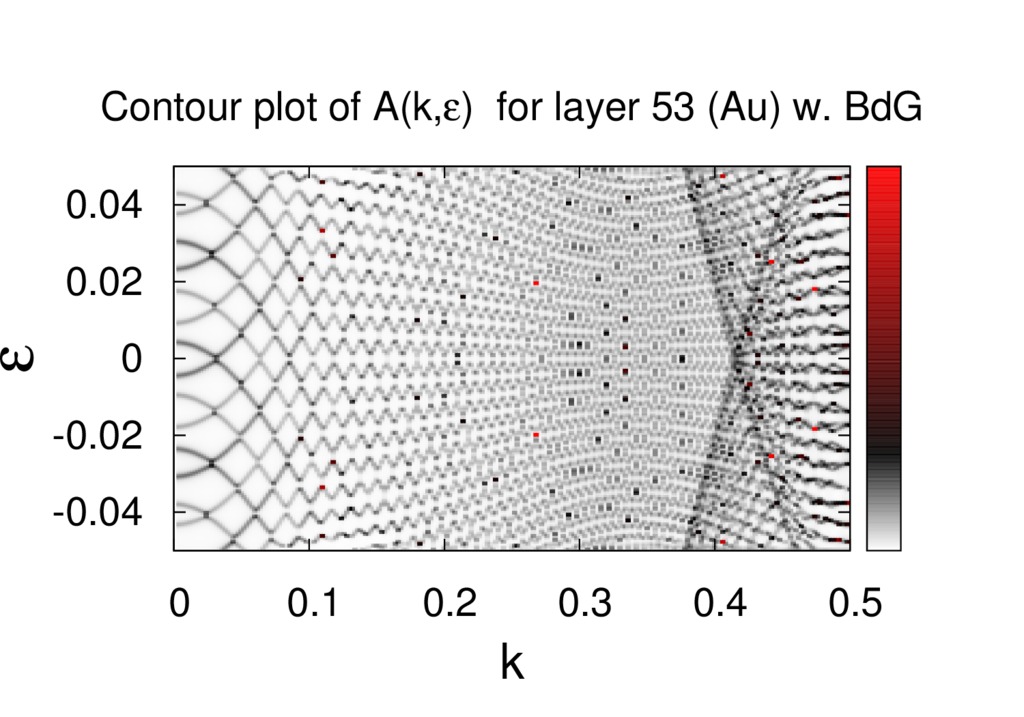}
   \caption{\label{fig:supra1}%
            (Color online)
            Contour plot of the BSF
            (quasiparticle spectrum calculated from BdG equations)
            from the "middle" of the Au layers for different thicknesses of the Au:
            3 Au layers (top left panel) and 9 Au layers (top right panel)
            24 Au layers (bottom left panel) and 93 Au layers (bottom right panel).  }
\end{figure}

We now consider the solution of the BdG equations described in the theory section, we model the pair potential in the inhomogeneous
Nb(100)/Au(100) system by assigning a constant value $\overline{\Delta}_{Nb}=0.05$~Ry to the Nb layers
and a constant $\overline{\Delta}_{Au}=0$~Ry for the Au layers (and the same to the empty sphere and vacuum layers).
The results of the calculation are shown in Fig.~\ref{fig:supra1}. Similarly to the normal state, first, we show results
for layers in the middle of the systems considered. However, we do not show any result for the bulk spectral function, 
because it is exactly zero in the energy range of the bulk superconducting gap.
What can be seen immediately is that there is a superconducting gap even in the Au layers. 
This gap must have been induced by the 
vicinity of the Nb, because  $\overline{\Delta}_{Au}=0$~Ry. Examining the details of the
quasiparticle spectrum, especially the one corresponding to the sample with 9 and 24 Au layers, reveals that not
only one, but in fact several gaps are opened.
This is in strong contrast to bulk superconductors, where the quasiparticle states can be obtained
from the electronic ones by mirroring them to the Fermi energy and opening up a gap. Our result modifies this picture so, that
 the proximity of a superconductor
in the studied heterostructures induces the mirroring of the electronic bands, and opens up a
gap -- which is significantly smaller than the one in the bulk -- at each band crossing.
This is valid for those band crossings as well that are not directly at the Fermi level
but within the $\overline{\Delta}_{Nb}$ energy range. In the case of the Nb/Au system
-- due to the the QW states in the normal state -- the result is a sort of oscillating quasi-band in Fig.~\ref{fig:supra1}.
This is a speciality of the Nb/Au system, other systems may not look so clean.
The induced gap is opened between the mirrored branches of the interface state as well.
However, in contrast to the QW states, the interface states do shift quite significantly  upwards in
energy, and these states still disappear rather quickly away from the interface.
 It should be noted as well that those regions of the spectrum which were more or less smeared out in the normal state,
now sharpened up. This is the consequence of the opening of the superconducting gap in the Nb: the states where scattering
into was allowed before, now disappeared.

In a quasiclassical picture one expect dispersionless ABS. 
However, this is not what the quasiparticle bands are showing.
What we find is, that the dispersion of the $k$-dependent ABS can be understood from the features of QW states
in the normal state as it was described above.
In  conventional superconductivity, the gap is assumed to be $k$-independent,
while in our calculations the obtained
energy gap strongly depends on the two-dimensional $k$. This is quite surprising, considering the fact that
our calculations involved only a totally conventional superconductivity scenario.
This is in an even larger contrast to the result of Suvasini et al.~\cite{Suvasini} who obtained only a very
week $k$-dependence of the gap in bulk Nb.
In this sense our results show similarities between the physics of conventional superconductor -- normal metal
heterostructures and unconventional superconductivity.

\begin{figure}[hbt!]
   \includegraphics[width=0.52\linewidth]{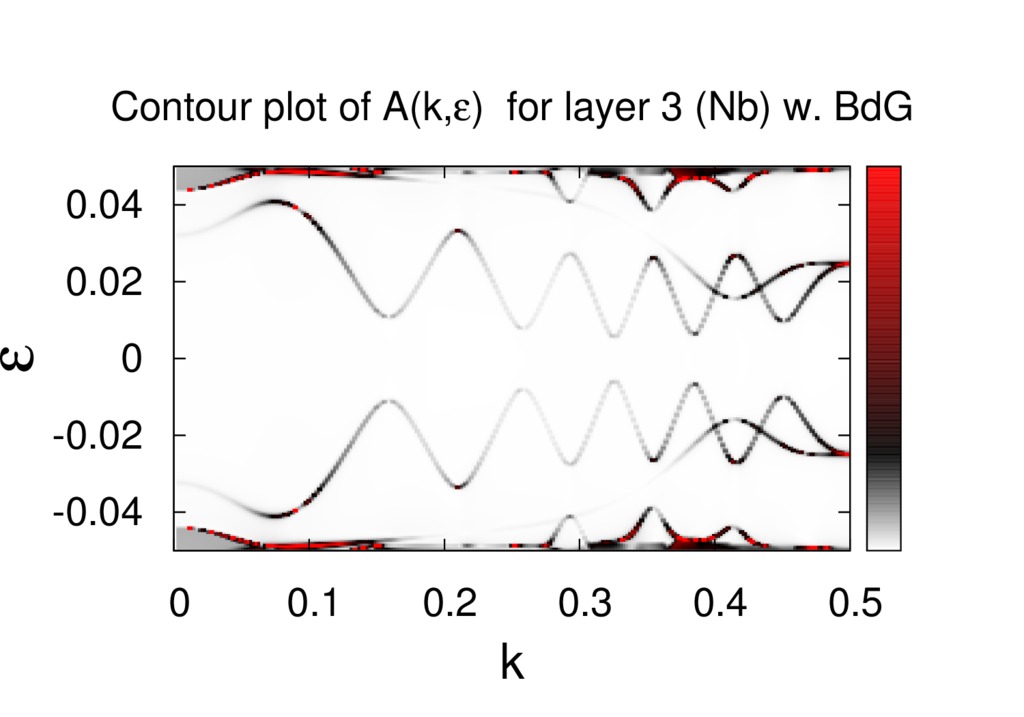}~
   \includegraphics[width=0.52\linewidth]{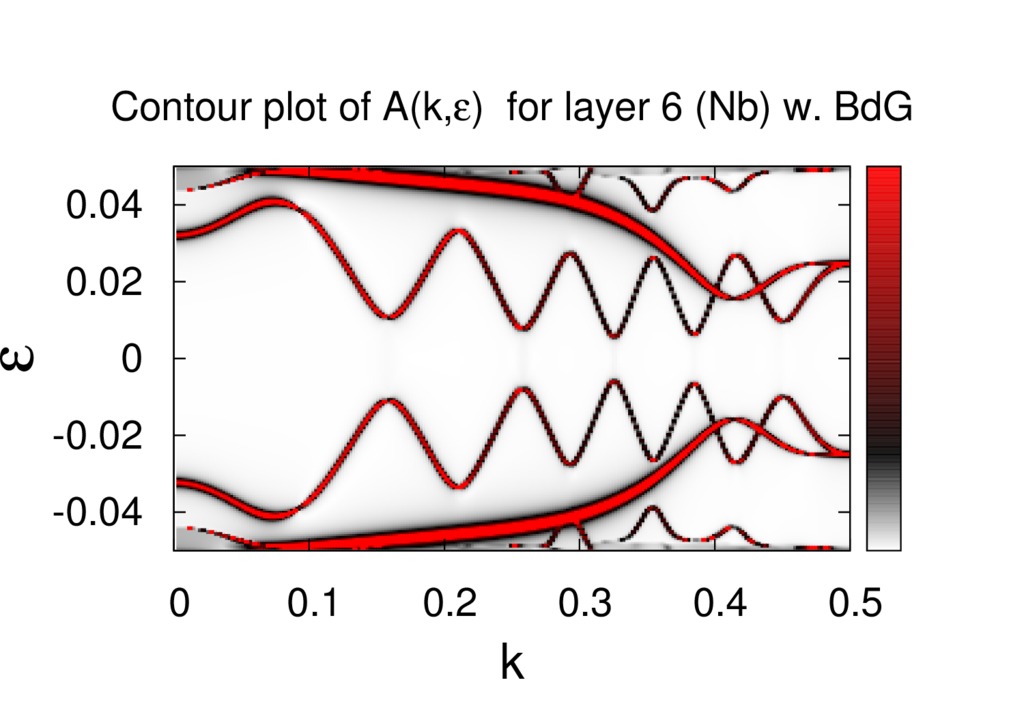}\\
   \includegraphics[width=0.52\linewidth]{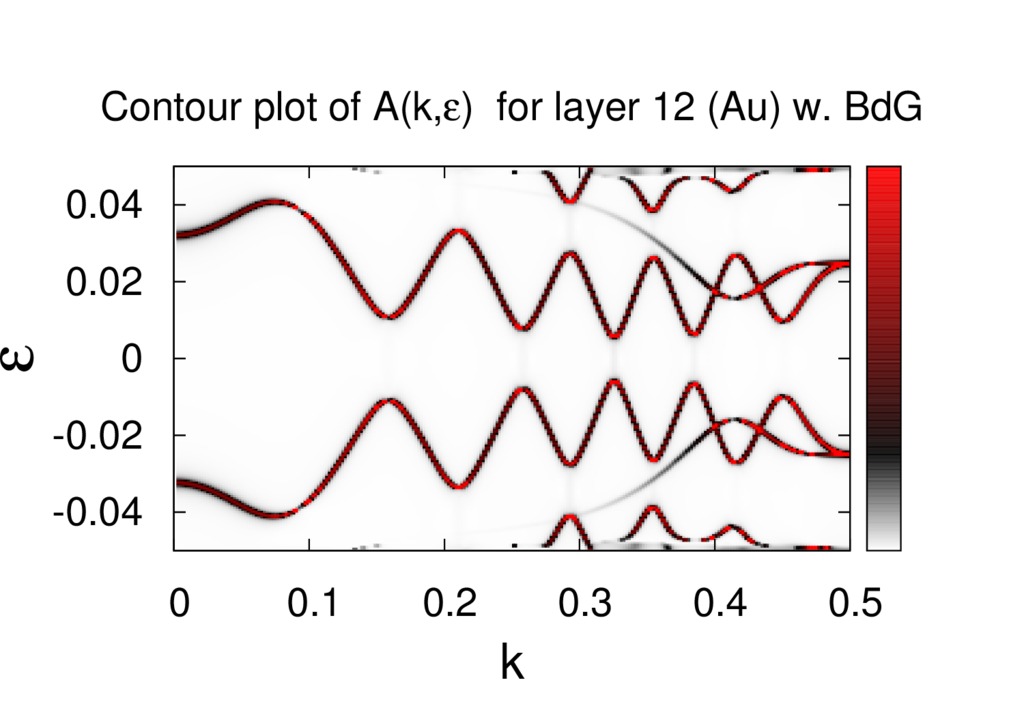}~
   \includegraphics[width=0.52\linewidth]{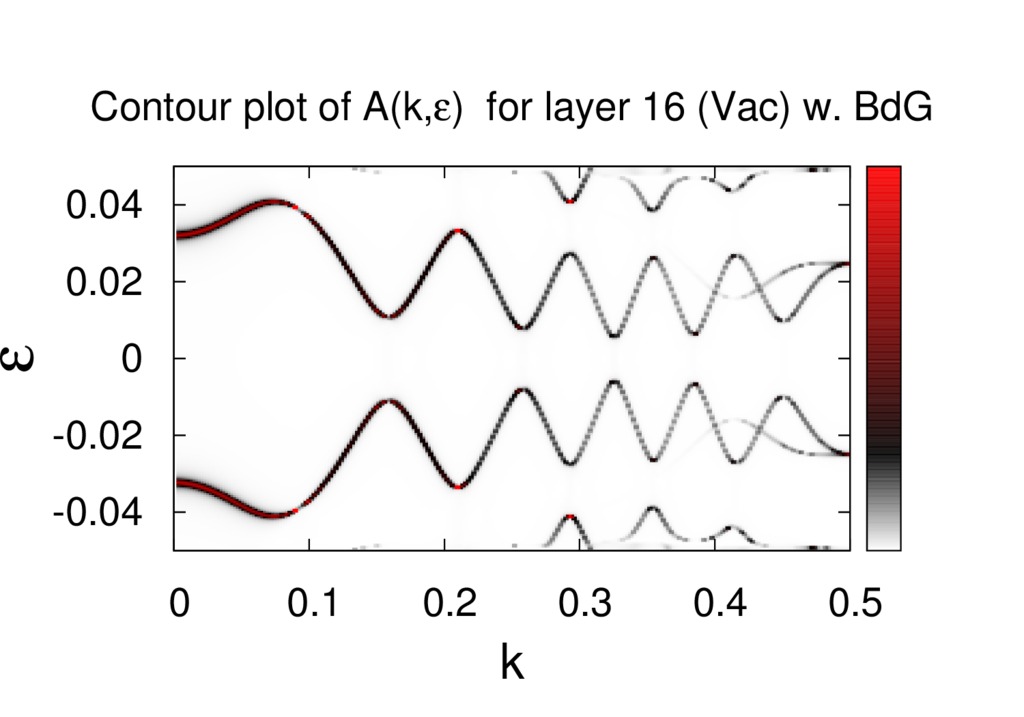}
   \caption{\label{fig:supra2}%
            (Color online)
            Contour plot of the BSF
            (quasiparticle spectrum calculated from BdG equations) for different layers in the system consisting of
            6 Nb layers, 9 Au layers and 3 empty sphere layers.}
\end{figure}

\begin{figure}[hbt!]
   \includegraphics[width=0.85\linewidth]{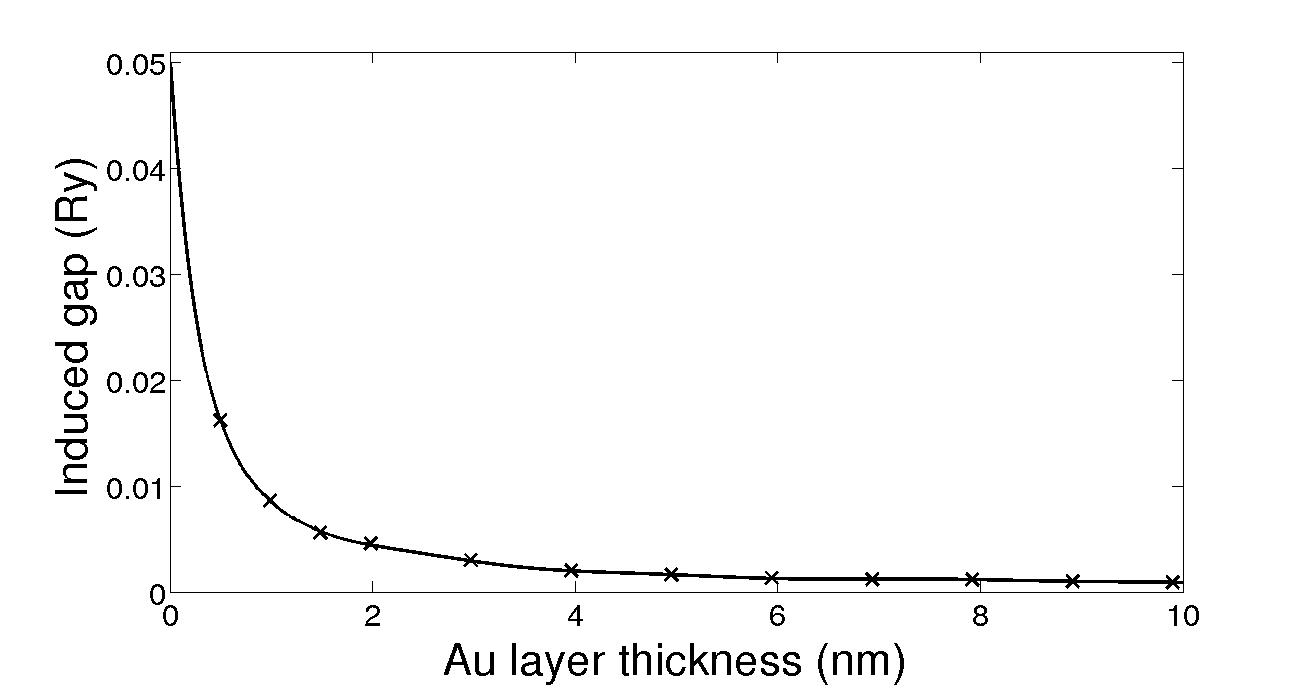}
   \caption{\label{fig:gap}%
             Induced gap on the Au layers as function of the thickness of the Au,
             extracted from results similar to that shown in Fig. \ref{fig:supra1}.}
\end{figure}

Further interesting features of the quasiparticle spectrum are revealed
if we analyze the spectrum layer by layer for a fixed system size (6 Nb and 9 Au layers, see Fig.~\ref{fig:supra2}).
As the QW states
did overlap with the Nb layers in the normal state, they still do in superconducting state.
However, as the quasiparticle states in the Au show a much
smaller gap than the one in the Nb, these overlapping states lessen the gap in the Nb layers next to the Au interface.
By performing further calculations, where the interfacial Nb layers were more numerous, we found that this effect decays quickly,
but can be observed up to 15 layers.
In the other side of the interface, in the Au layers the induced gap remains constant for each layer.
Therefore, an induced superconductivity may  be observed in the Au overlayers.
This is in accord of the experimental observations, where it was found that
the whole Nb/Au system is superconducting
(a common $T_c$ has been obtained experimentally in Refs.~\onlinecite{Yamazaki1, Yamazaki2}). 
Cooper pairs can be found in the whole system,
and the induced gap -- that appeared in the quasiparticle spectrum in each of the Au layers -- can be interpreted as a 
consequence of an effective electron-phonon coupling in the Au overlayers caused by the semi-infinite Nb.
Quite surprisingly, in our calculations we did not find any layer dependence of the induced gap.
This can be attributed to the fact that we did not consider the layer dependence of the pairing potential, or by other 
words, the layer dependence of the electron-phonon interaction was neglected.
Nevertheless, the size of the gap does change with the thickness of the system, as it can be seen in 
Fig.~\ref{fig:supra2} and also summarized in Fig.~\ref{fig:gap}. It shows a fast decay, however, it can not be fitted well
by an exponential function. 

It is also useful to mention that for $k=0$ the spectrum of the ABS is
comparable with the results of one-dimensional model calculations
and the Andreev energy levels show the similar $1/L$ dependence
which were also obtained in Ref.~\onlinecite{Cserti1}.
However, we emphasize that this property is the consequence of the roughly $2\pi/L$ sampling connected to the QW states
and can not be regarded as an universal feature for every S/N heterostructures.

\subsection{Surface states}

Metallic surfaces often exhibit a Shockley-type surface
state. The energy of such states are located in a relative band gap of the bulk, normal state band
structure and usually have a parabolic dispersion, and therefore such electrons behave like a nearly
free two dimensional electron gas.
Surface states are easily accessible to spectroscopy with photoemission, since they are often located near the Fermi energy.
Therefore, it is interesting to study such surface states once the material becomes superconducting.

\begin{figure}[hbt!]
   \includegraphics[width=0.52\linewidth]{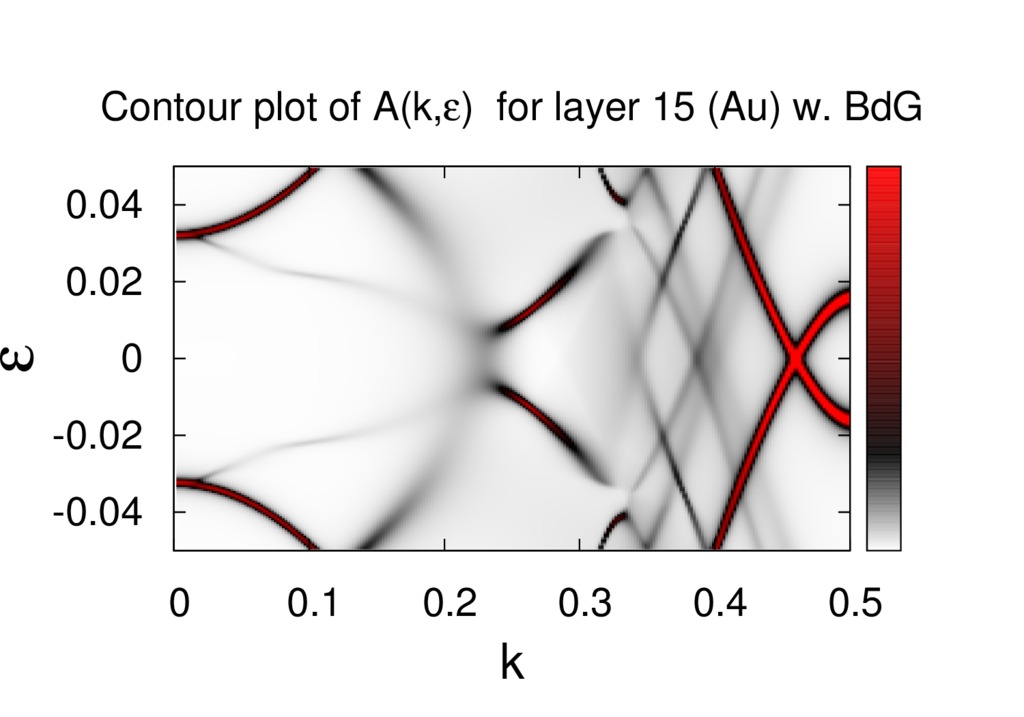}~
   \includegraphics[width=0.52\linewidth]{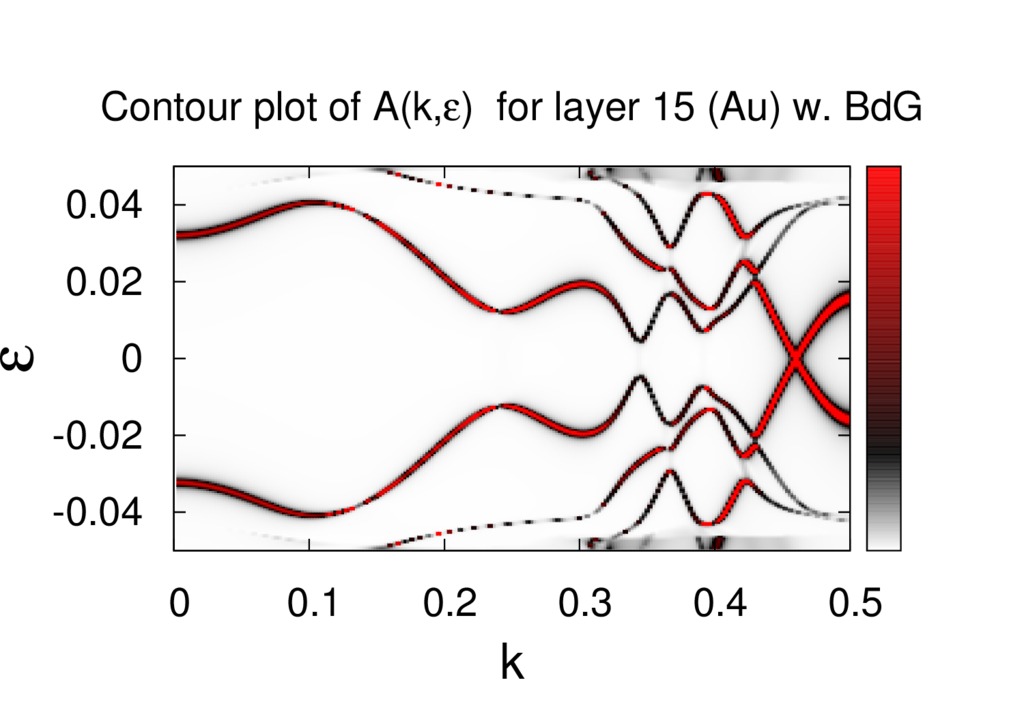}
   \caption{\label{fig:surf}%
            (Color online) Contour plot of the BSF in the $k_y=0$ direction corresponding the last layer of Au.
            The Au sample consisted 9 layers.
            The quasiparticle spectrum was calculated from BdG equations.
            $\overline{\Delta}_{Nb}=0$~Ry is used on the left panel and $\overline{\Delta}_{Nb}=0.05$~Ry on the right panel.
}
\end{figure}

We calculated such Shockley-type surface state from the BdG equations in the case of the investigated
Nb/Au heterostructure along the direction $k_y =0$. It should be emphasized that this surface state is entirely 
fictitious, as this surface is of an Au(100) in the BCC lattice structure. First, setting the $\overline{\Delta}_{Nb}=0$~Ry
(see the left panel of Fig.~\ref{fig:surf}), the surface state can be observed in the normal state electronic structure.
While applying a finite $\overline{\Delta}_{Nb}$
pairing potential does open up a gap in the Au, just as we discussed earlier for the case of $k_x=k_y$ direction,
but no gap opens at the crossing of the surface state bands, indicating that it does not couple to the superconductor.
This effect can be attributed to the fact that obviously the surface state is quite localized to the
top layers of the metal surface and it is mainly isolated from
the bulk states.
Consequently, they do not take part in the Andreev scattering process and thus they do not have a gap in the spectrum,
as it can be seen in Fig.~\ref{fig:surf}.

As we indicated earlier, an opposite behavior could be observed for the interface state,
which is localized to the Nb/Au interface.
The energy of these states shifts upwards. This  can be explained
by the stronger interaction between the superconductor and the normal metal
resulted in a larger gap than in the QW states. 

\section{Summary}

In this paper we have presented the first material specific calculations
for an s-wave superconductor -- normal metal heterostructure.
Based on first principles BdG equations a novel and
unique computer code was developed which allows us to study the nature of the Andreev bound states
related to the proximity effect in normal metal -- superconductor heterostructures.

For the first time we have extended the SKKR method for the solution of the
KSBdG Eqs.~(\ref{eq:KSBdG}) which allows one to investigate the quasiparticle spectrum of superconducting heterostructures.
In order to compare our results with normal state electronic structure calculations,
a scalar relativistic generalization of the BdG equations within Multiple Scattering Theory was also provided.
Formally, the generalized Faulkner-Stocks formula, given by Eq.~(\ref{eq:genFS}), is the main result of this paper.

To illustrate the power of the new method, it was applied to Nb/Au heterostructures.
For simplicity, Au overlayers of BCC(100) lattice structure on a Nb BCC(100) host have been investigated.
While such material is not likely to exist for larger
Au thicknesses, by assuming a layer by layer growth, it resulted in an easily understandable system with quantum well states.
The effect of the superconducting host on the quasiparticle spectrum of Au overlayers 
can be more easily identified by these states than on a more complex band structure of a real material.
Calculations for a more realistic geometry will be published elsewhere.

We showed that the QW states (we found to exist in the normal state band structure calculations)
become bound Andreev states due to Andreev scattering.
The major result of our investigations is that the ABS have dispersion,
which can be obtained only by developing the BdG-SKKR method.
We also found that the proximity of a superconductor
in the studied heterostructures induces the mirroring of the electronic bands, and opens up a
gap at each band crossing, and the gaps are strongly $k$-dependent.
We have seen that this induced gap remains constant for each layer for a given Au thickness,
however, the size of the gap decays as function of the Au thickness.
For $k=0$, the one-dimensional model calculations of the Andreev energy levels~\cite{Cserti1} are
recovered for those heterostructures where the nearly free electron
approach is applicable.
We also investigated the properties of the
surface state at the Au surface and found that the gap does not appear in
the energy spectrum of these states, probably, because they
are localized to the surface and consequently do not take part in the Andreev scattering
process.

{\emph{Acknowledgment}} ---  This work was supported by the Hungarian Science Foundation
OTKA under the contracts No. K108676 and K109570.
The authors would like to thank L{\'a}szl{\'o} Szunyogh and Istv{\'a}n T{\"u}tt{\H o} for helpful discussions.
G{\'a}bor Csire thanks ESF for supporting a visit to the University of Bristol.

\begin{widetext}
\appendix*
\section{Mathcing the wavefunctions \label{appen}}
The boundary conditions, at the muffin-tin sphere boundary, can be expressed
as follows ($a=e,h$) for the radial part of the wavefunction:
\begin{equation}
 r_{mt} R_l^{a}(r=r_{mt}) = A_l^a~ P_l^{a,(1)}(x=x_{mt}) + B_l^a~ P_l^{a,(2)}(x=x_{mt}),
\end{equation}
\begin{equation}
 r_{mt} \left. \frac{\dd}{\dd r} \left( r R_l^{a} (r) \right) \right|_{r=r_{mt}} =
 A_l^a~ \left. \frac{\dd}{\dd x}P_l^{a,(1)}(x) \right|_{x=x_{mt}}
 + B_l^a~ \left. \frac{\dd}{\dd x}P_l^{a,(2)}(x) \right|_{x=x_{mt}},
\end{equation}
where $P_l^{a,(1)}(x)$ and $P_l^{a,(2)}(x)$ are the regular solutions
of the scalar relativistic BdG Eqs.~(\ref{eq:SRBdG}) inside the muffin-tin sphere, and $r_{mt}$ is the
radius of the muffin-tin sphere ($x_{mt}=\log r_{mt}$).
We emphasize that there are two independent regular and two independent irregular solutions of Eqs.~(\ref{eq:SRBdG}).

The matching conditions can be written in matrix form:
\begin{equation}
 \mathbf{M} \underline{a}^e= \underline{b}^e,
\qquad
 \mathbf{M} \underline{a}^h= \underline{b}^h,
\end{equation}
where
\begin{equation}
 \underline{a}^e=\begin{pmatrix}
     A_l^e \\ B_l^e \\ {t}_{l}^{ee} \\ t_{l}^{he}
                       \end{pmatrix},
\qquad
\underline{a}^h=\begin{pmatrix}
     A_l^h \\ B_l^h \\ {t}_{l}^{eh} \\ t_{l}^{hh}
                       \end{pmatrix},
\end{equation}
\begin{equation}
\underline{b}^e=\begin{pmatrix}
    r_{mt} j_l(p^e r_{mt}) \\ 0 \\ r_{mt} j_l(p^e r_{mt}) + r_{mt}^2 p^e j'_l(p^e r_{mt}) \\ 0
                       \end{pmatrix},
\qquad
\underline{b}^h=\begin{pmatrix}
    0 \\ r_{mt} j_l(p^h r_{mt}) \\ 0 \\ r_{mt} j_l(p^h r_{mt}) + r_{mt}^2 p^h j'_l(p^h r_{mt})
                       \end{pmatrix},
\end{equation}
\begin{equation}
\mathbf{M}=
  \begin{pmatrix}
    P^{e,(1)}_l(x_{mt}) & P^{e,(2)}_l(x_{mt}) & \ii p^e r_{mt} h_l^+(p^e r_{mt}) & 0\\
    P^{h,(1)}_l(x_{mt}) & P^{h,(2)}_l(x_{mt}) & 0 & -\ii p^h r_{mt} h_l^+(p^h r_{mt}) \\
    \partial_x P^{e,(1)}_l(x_{mt}) & \partial_x  P^{e,(2)}_l(x_{mt}) & \ii r_{mt} p^e (1+p^e r_{mt}\partial_r) h_l^+(p^e r_{mt}) & 0 \\
    \partial_x P^{h,(1)}_l(x_{mt}) & \partial_x  P^{h,(2)}_l(x_{mt}) & 0 & -\ii r_{mt} p^h (1+p^h r_{mt}\partial_r) h_l^+(p^h r_{mt})
  \end{pmatrix}.
\end{equation}

The regular wavefunctions can be continued inside the muffin-tin sphere as follows
\begin{itemize}
\item for electron-like incoming wave
\begin{equation}
      r \begin{pmatrix}
           R_l^{ee}(r)\\
           R_l^{he}(r)
      \end{pmatrix}
           =
       r \begin{pmatrix}
           j_l(p^e r) - \ii p^e t_{l}^{ee} h_l^+(p^e r)\\
           \ii p^h t_{l}^{he} h_l^+(p^h r)
      \end{pmatrix}
      \rightarrow        A_l^e
              \begin{pmatrix}
               P_l^{e,(1)}(r) \\ P_l^{h,(1)}(r)
              \end{pmatrix} +
               B_l^e
              \begin{pmatrix}
               P_l^{e,(2)}(r) \\ P_l^{h,(2)}(r)
              \end{pmatrix},
\end{equation}
\item for hole-like incoming wave
\begin{equation}
      r \begin{pmatrix}
           R_l^{eh}(r)\\
           R_l^{hh}(r)
      \end{pmatrix}
      =
       r \begin{pmatrix}
           - \ii p^e t_{l}^{eh} h_l^+(p^e r)\\
           j_l(p^h r) + \ii p^h t_{l}^{hh} h_l^+(p^h r)
      \end{pmatrix}
            \rightarrow        A_l^h
              \begin{pmatrix}
               P_l^{e,(1)}(r) \\ P_l^{h,(1)}(r)
              \end{pmatrix} +
               B_l^h
              \begin{pmatrix}
               P_l^{e,(2)}(r) \\ P_l^{h,(2)}(r)
              \end{pmatrix},
\end{equation}
\end{itemize}
and the irregular solutions as
\begin{equation}
      r
      \begin{pmatrix}
           J_l^{ee}(r) & 0 \\
           0        & J_l^{hh}(r)
      \end{pmatrix}
       \rightarrow
             \begin{pmatrix}
           I_l^{ee}(r) & I_l^{eh}(r) \\
           I_l^{he}(r)  & I_l^{hh}(r)
      \end{pmatrix}.
      \label{eq:irreg_match}
\end{equation}

Also, to calculate the Green function, given by Eq. (\ref{eq:genFS}), 
the determination of the normalized irregular solution, inside the muffin-tin sphere, is indispensable.
To satisfy the matching conditions, one needs to use
the linear combination of the regular solutions, $P_l^{a,(1)}(x)$, $P_l^{a,(2)}(x)$, and the
irregular solutions, $\widetilde P_l^{a,(1)}(x)$, $\widetilde P_l^{a,(2)}(x)$:
\begin{equation}
   \begin{pmatrix}
     I_l^{ee}(x) \\
     I_l^{he}(x)
   \end{pmatrix}
     =\tilde A_l^e \begin{pmatrix} P_l^{e,(1)}(x) \\ P_l^{h,(1)}(x) \end{pmatrix} +
      \tilde B_l^e \begin{pmatrix} P_l^{e,(2)}(x) \\ P_l^{h,(2)}(x) \end{pmatrix} +
      \tilde C_l^e \begin{pmatrix} \widetilde P_l^{e,(1)}(x) \\ \widetilde P_l^{h,(1)}(x) \end{pmatrix} +
      \tilde D_l^e \begin{pmatrix} \widetilde P_l^{e,(2)}(x) \\ \widetilde P_l^{h,(2)}(x) \end{pmatrix} ,
\end{equation}
\begin{equation}
   \begin{pmatrix}
     I_l^{eh}(x) \\
     I_l^{hh}(x)
   \end{pmatrix}
     =\tilde A_l^h \begin{pmatrix} P_l^{e,(1)}(x) \\ P_l^{h,(1)}(x) \end{pmatrix} +
      \tilde B_l^h \begin{pmatrix} P_l^{e,(2)}(x) \\ P_l^{h,(2)}(x) \end{pmatrix} +
      \tilde C_l^h \begin{pmatrix} \widetilde P_l^{e,(1)}(x) \\ \widetilde P_l^{h,(1)}(x) \end{pmatrix} +
      \tilde D_l^h \begin{pmatrix} \widetilde P_l^{e,(2)}(x) \\ \widetilde P_l^{h,(2)}(x) \end{pmatrix} ,
\end{equation}
where
\begin{equation}
    \begin{pmatrix}
     \tilde A_l^e \\ \tilde B_l^e \\ \tilde C_l^e \\ \tilde D_l^e
  \end{pmatrix}=
    \mathbf N^{-1}
    \begin{pmatrix}
     r_{mt} j_l(p^e r_{mt})\\
     0\\
     r_{mt} \left(1 + p^e r_{mt} \partial_r \right) j_l(p^e r_{mt}) \\
     0
  \end{pmatrix},
\end{equation}
\begin{equation}
    \begin{pmatrix}
     \tilde A_l^h \\ \tilde B_l^h \\ \tilde C_l^h \\ \tilde D_l^h
    \end{pmatrix}=
    \mathbf N^{-1}
    \begin{pmatrix}
    0\\
    r_{mt} j_l(p^h r_{mt})\\
    0 \\
    r_{mt} \left( 1 + p^h r_{mt} \partial_r \right) j_l(p^h r_{mt})
  \end{pmatrix},
\end{equation}
\begin{equation}
  \mathbf N =
 \begin{pmatrix}
   P^{e,(1)}_l(x_{mt}) & P^{e,(2)}_l(x_{mt}) & \widetilde P^{e,(1)}_l(x_{mt}) & \widetilde P^{e,(2)}_l(x_{mt})\\
    P^{h,(1)}_l(x_{mt}) & P^{h,(2)}_l(x_{mt}) & \widetilde P^{h,(1)}_l(x_{mt}) & \widetilde P^{h,(2)}_l(x_{mt}) \\
    \partial_x P^{e,(1)}_l(x_{mt}) & \partial_x P^{e,(2)}_l(x_{mt}) &
    \partial_x \widetilde P^{e,(1)}_l(x_{mt}) & \partial_x \widetilde P^{e,(2)}_l(x_{mt}) \\
    \partial_x P^{h,(1)}_l(x_{mt}) & \partial_x  P^{h,(2)}_l(x_{mt}) &
    \partial_x \widetilde P^{h,(1)}_l(x_{mt}) & \partial_x \widetilde P^{h,(2)}_l(x_{mt})
 \end{pmatrix}.
 \end{equation}
\end{widetext}

\bibliographystyle{apsrev4-1}
\bibliography{bdg_skkr}

\end{document}